\documentclass[12pt,a4paper]{article}

\usepackage{jheppub}
\usepackage[utf8]{inputenc}
\usepackage[english]{babel}
\usepackage{color}
\usepackage{amsmath}
\usepackage{verbatim}   
\usepackage{bbm}
\usepackage{subfigure}  
\usepackage{amsfonts}
\usepackage{amssymb}
\usepackage{mathrsfs}
\usepackage{stmaryrd}
\SetSymbolFont{stmry}{bold}{U}{stmry}{m}{n}
\usepackage{graphicx}
\usepackage{slashed}
\usepackage{amsthm}
\usepackage{graphicx}
\usepackage{dcolumn}
\usepackage{bm}
\usepackage{color}
\usepackage{amsmath}
\usepackage{verbatim}   
\usepackage{subfigure}  
\usepackage{amsfonts}
\usepackage{amssymb}
\usepackage{mathrsfs}
\usepackage{slashed}
\usepackage{amsthm}
\usepackage{bbold}
\usepackage{cleveref}
\usepackage[]{units} 
\usepackage{color}
\usepackage{units}
\usepackage{mathtools}

\def\d{\text{d}}

\makeatletter   
\renewcommand*\env@matrix[1][*\c@MaxMatrixCols c]{%
  \hskip -\arraycolsep
  \let\@ifnextchar\new@ifnextchar
  \array{#1}}
\makeatother

\newcommand{\be}{\begin{equation}}
\newcommand{\ee}{\end{equation}}
\newcommand{\bea}{\begin{eqnarray}}
\newcommand{\eea}{\end{eqnarray}}

\newcommand{\cF}{\mathcal{F}}

\newcommand{\cM}{\mathcal{M}}

\begin{document}

\title{Threshold corrections in heterotic flux compactifications}
\author[a]{Carlo Angelantonj}
\author[b,c]{Dan Isra\"el}
\author[b,c]{Matthieu Sarkis}

\affiliation[a]{Dipartimento di Fisica, Universit\`a di Torino, and INFN Sezione di Torino
          \\
            Via P. Giuria 1, 10125 Torino, Italy}
\affiliation[b]{LPTHE, UMR 7589, Sorbonne Universit\'es, UPMC Univ. Paris 06, 4 place Jussieu, Paris, France}
\affiliation[c]{CNRS, UMR 7589, LPTHE, F-75005, Paris, France}

\emailAdd{carlo.angelantonj@unito.it}
\emailAdd{israel@lpthe.jussieu.fr}
\emailAdd{msarkis@lpthe.jussieu.fr}

\null\vskip10pt
\abstract{We compute the one-loop threshold corrections to the gauge and gravitational couplings for a large class of 
$\mathcal{N}=2$ non-K\"ahler heterotic compactifications with three-form flux, 
consisting in principal two-torus bundles over $K3$ surfaces. We obtain the results as sums of BPS-states contributions, depending 
on the topological data of the bundle. We analyse also the worldsheet non-perturbative 
corrections coming from instantons wrapping the torus fiber, that are mapped under S-duality to D-instanton corrections 
in type I flux compactifications.}
                              
\maketitle


\section{Introduction}

Supersymmetric heterotic compactifications constitute one of the main approaches to particle physics phenomenology from 
sting theory. The conditions  ensuring at least $\mathcal{N}=1$ supersymmetry in spacetime, at order $\alpha'$, 
are encoded in the Hull-Strominger system, a set of BPS equations constraining the internal geometry~\cite{Hull:1986kz,Strominger:1986uh}. A well-known class of solutions to this system consists of a Calabi-Yau 3-fold equipped with a stable holomorphic vector bundle. This type of construction leads to GUT groups which can be smaller than $E_6$, but also come with a collection of moduli which are undesirable from a phenomenological point of view.

To tackle this moduli problem, one can consider non-K\"ahler compactifications with non-trivial fluxes for the  Kalb-Ramond three-form 
field strength along the internal geometry. The high level of complexity of the Hull-Strominger system in this general setting forbids a generic discussion of its solutions. One large family of flux compactifications, originally obtained in~\cite{Dasgupta:1999ss} from 
string dualities, and often denoted in the literature as Fu-Yau geometry, has however been studied quite 
extensively, see $e.g.$~\cite{Goldstein:2002pg,Becker:2006et,2008arXiv0807.0827B}. The internal manifold consists of a 
principal two-torus bundle over a warped $K3$ surface, equipped with the pullback of a stable holomorphic vector 
bundle over the base. One may also possibly add an Abelian bundle over the total space which reduces to Wilson lines 
in the more familiar $K3\times T^2$ setting, however we will not consider them in this paper. 

A subfamily of these non-K\"ahler solutions, leading to $\mathcal{N}=2$ supersymmetry in spacetime, has been shown to 
be amenable to a gauged linear sigma model description~\cite{Adams:2006kb} on the heterotic string worldsheet. 
This approach allows in principle to extract the massless spectrum using Landau-Ginzburg cohomological methods \cite{2011JHEP...03..045A} 
and was used to prove T-duality symmetries in these curved flux backgrounds~\cite{2013JHEP...11..093I}.

Localization techniques were then used by two of the present authors to compute the new supersymmetric index~\cite{Israel:2015aea,Israel:2016xfu}, which is, in the context of heterotic $\mathcal{N}=2$ compactifications, the building block 
that is used  to compute the threshold corrections to the BPS-saturated couplings in the
 low energy four-dimensional $\mathcal{N}=2$ supergravity action~\cite{1996NuPhB.463..315H}.

The goal of this paper is precisely to compute explicitely the threshold corrections to the gravitational and gauge couplings of these  compactifications with torsion, thus extending the results already known for $K3\times T^2$ compactifications or orbifolds 
thereof~\cite{1996NuPhB.482..187H,Kiritsis:1996dn,Stieberger:1998yi,Datta:2015hza,Angelantonj:2014dia,Mayr:1993mq,Antoniadis:1993au,Kaplunovsky:1992vs,LopesCardoso:1994ik,Mayr:1995rx,Gregori:1997hi} and results for 
local models of non-K\"ahler compactifications~\cite{Carlevaro:2012rz}. The threshold corrections are written 
naturally as the integral of some almost holomorphic modular form over the fundamental domain of the worldsheet modular group.

This type of integral can be computed using the standard orbit method that was developped for $K3 \times T^2$ compactifications, 
which consists in unfolding the integration domain against the Narain lattice partition 
function~\cite{Kaplunovsky:1987rp}. This approach is convenient for studying the 
D-instanton corrections in the type I S-duals (see $e.g.$~\cite{Bachas:1997mc,Camara:2008zk}), however it hides the explicit 
covariance under the perturbative duality group $O(2,2;\mathbbm{Z})$ of the two-torus, that occurs also in the 
$\mathcal{N}=2$ compactifications with torsion under study~\cite{2013JHEP...11..093I}.

Another approach, developed recently in~\cite{Angelantonj:2011br,Angelantonj:2012gw,Angelantonj:2015rxa} , suggests to maintain 
the explicit covariance under T-duality by instead keeping the Narain partition function intact,  expanding 
the remaining weak almost holomorphic modular form in terms of (absolutely convergent) Niebur-Poincar\'e series, and finally 
unfolding the integration domain against the latter. This approach not only has the advantage of keeping 
T-duality manifest and the analytic structure of the amplitude transparent, but rather it is the best 
(if not the only) way to extract physical couplings for values of the moduli close to the string scale, 
where the conventional expansion might fail to converge. This is especially useful for the 
present class of models, given that the volume of the two-torus fiber is generically frozen by the fluxes 
to a small value in string units. 

Following this approach, we obtain in this work compact and T-duality covariant expressions for the 
threshold corrections, written in a chamber-independent form, $i.e.$ valid for any values of 
the moduli of the torus fiber. The results depend explicitely on the topology of the principal 
two-torus bundle, $i.e.$ on the choice of a pair of anti-self-dual $(1,1)$ forms on the $K3$ base. 

We will consider thereafter an alternative representation of the threshold corrections in terms of a 
Fourier series expansion in the K\"ahler modulus $T$ of the torus fiber~\cite{Angelantonj:2012gw}, enlightening the origin of the 
various contributions, especially those corresponding to the worldsheet instantons wrapping the $T^2$.  
These corrections, that would be, for $\text{Spin}(32)/\mathbbm{Z}_2$ compactifications, 
S-dual to D1-instanton corrections in type I compactifications 
with Ramond-Ramond fluxes, are particularly interesting. Indeed, topologically, the two-torus is not a proper 
two-cycle of the total space of the bundle, but only a torsion two-cycle. Nevertheless as we will find the 
instanton corrections take the form of a infinite sum over the wrapping number. 

\paragraph{Conventions:}$T$ and $U$ denote respectively the complexified K\"ahler and complex structure moduli 
of the torus fiber. $\text{d}\nu=\text{d}\tau_1\text{d}\tau_2/\tau_2^2$ denotes the Poincar\'e measure on the complex 
upper-helf plane $\mathbbm H$. $\theta(\tau,z)$ is the odd Jacobi theta function. For definiteness we consider the $E_8\times E_8$ 
ten-dimensional heterotic string theory unless otherwise stated. 


\section{\texorpdfstring{$\mathcal{N}=2$}{N=2} heterotic threshold corrections and the new supersymmetric index}

We will be interested in a class of non-K\"ahler heterotic 
compactifications to four-dimensions,  corresponding to a principal bundle 
$T^2\hookrightarrow\mathcal{M}\stackrel{\pi}{\to}\mathcal{S}$ over a warped K3 surface 
$\mathcal{S}$ with three-form flux. 

These compactifications preserve $\mathcal{N}=2$ supersymmetry in space-time hence one can compute, as in the 
case of ordinary $K3 \times T^2$ compactifications, the one-loop corrections to the couplings of some 
two-derivative BPS-saturated terms in the four-dimensional low energy effective supergravity action. 
We will focus on the gravitational coupling and the gauge couplings associated with the different 
factors of the spacetime gauge group left unbroken by the choice of gauge bundle. 

The one-loop running of the coupling constant associated with a simple factor $G$ of the space-time gauge group is 
expressed through the relation:
\begin{equation}
\label{eq:running}
\frac{16\pi^2}{g_G^2(\mu)}=\frac{16\pi^2}{g_s^2}+\beta_G\log\frac{M_s^2}{\mu^2}+\Delta_G\,.
\end{equation}
The second term in the right-hand-side of \cref{eq:running} corresponds to the contribution from the massless 
multiplets, hence to the running one would compute 
in a field theoretic setting. It is proportionnal to the gauge-theory beta-function $\beta_G$. The 
last term $\Delta_G$ incorporates the contribution from the whole tower of massive fields, 
hence describes the stringy part of the one-loop correction to gauge coupling. 

A similar expression holds for the one-loop threshold correction to the  gravitational coupling as well:
\begin{equation}
\frac{16\pi^2}{g_\text{grav}^2(\mu)}=\frac{16\pi^2}{g_s^2}+\beta_\text{grav}\log\frac{M_s^2}{\mu^2}+\Delta_\text{grav}\,.
\end{equation}

These threshold corrections have been studied in great details for $K3 \times T^2$ compactifications, see 
the introduction for a partial 
list of relevant references. Extended $\mathcal{N}=2$ supersymmetry in spacetime highly 
constrains these corrections; in particular, they only receive contribution 
from BPS states. It turns out that they all can be expressed as the integral over the 
fundamental domain of the worldsheet torus modular group of descendants of 
a quantity known as the new supersymmetric index~\cite{1992NuPhB.386..405C}. 
This objet is independent of the moduli of the K3 surface, but depends on the 
torus and Wilson line moduli of the compactification. 

This new supersymmetric index is defined by the following trace in the Ramond sector of the right-moving fermions:
\begin{equation}
\label{eq:susy_index_def}
Z_\textsc{new}(\tau,\bar\tau)=\frac{1}{\eta(\tau)^2}\,\text{Tr}_\textsc{r}\left(\bar{J}_{\, 0}
(-1)^{F_\textsc{r}} q^{L_0-c/24}\bar{q}^{\bar{L}_0-\bar{c}/24}\right)\,.
\end{equation}
It was shown by Harvey and Moore \cite{1996NuPhB.463..315H} that this new 
supersymmetric index counts the BPS multiplets in spacetime, since 
worldsheet supersymmetry dictates that :
\begin{equation}
\label{eq:BPSsum}
-\frac{1}{2i \eta^2}Z_\textsc{new} (q,\bar q) = \sum_{\text{BPS vectors}}q^\Delta \bar{q}^{\bar \Delta}-\sum_{\text{BPS hypers}} 
q^\Delta \bar{q}^{\bar \Delta}\, .
\end{equation}
For the $\mathcal{N}=2$ torsional compactifications of interest, 
the new supersymmetric index was computed in~\cite{Israel:2015aea}, using a purposely designed 
gauged linear sigma model~\cite{Adams:2006kb} and supersymmetric localization of the path 
integral~\cite{2015CMaPh.333.1241B}.

A main difference with the conventional $K3 \times T^2$ compactifications is that 
both the complex structure and complexified K\"ahler 
moduli of the two-torus fiber are now generically quantized. However Abelian bundles over the 
total space, that would reduce to Wilson lines in the $K3\times T^2$ case, have moduli which are not quantized 
by the three-form flux; the new supersymmetric index as function of these moduli was computed~\cite{Israel:2016xfu}. 
For simplicity we will assume that these extra moduli are turned off. 

These compactifications are also characterized by the pullback of a holomorphic vector bundle 
$\mathcal{V}$ over the  $K3$ base. For definiteness, we will embed its structure group in the first $E_8$ factor of the 
$E_8\times E_8$ heterotic gauge group.

Then the new supersymmetric index $Z_\textsc{new}$, which was computed in \cite{Israel:2015aea}, is expressed in terms of a 
non-holomorphic dressed elliptic genus $Z_\textsc{fy}\left(\tau,\bar{\tau},z\right)$  through
\begin{equation}
Z_\textsc{new}(\tau,\bar\tau)=\frac{\bar{\eta}^{2}E_{4}(\tau)}{2\eta^{10}}\sum_{\gamma,\delta=0}^{1}q^{\gamma^{2}}
\left.\left\{\left(\frac{\theta\left(\tau,z\right)}{\eta(\tau)}\right)^{8-r}
Z_\textsc{fy}\left(\tau,\bar{\tau},z\right)\right\}\right|_{z=\frac{\gamma\tau+\delta}{2}}\, ,\label{eq:index}
\end{equation}
where we have defined the non-holomorphic dressed elliptic genus as follows:
\begin{equation}
\label{eq:def_dressed}
Z_\textsc{fy}\left(\tau,\bar{\tau},z\right) = \frac{1}{\bar{\eta}(\bar{\tau})^{2}}\,
\text{Tr}_{\textsc{rr},\mathcal{H}_\textsc{fy}}\left(e^{2i\pi z J_{0}}\bar{J}_{\, 0}
(-1)^{F} q^{L_0-c/24}\bar{q}^{\bar{L}_0-\bar{c}/24} \right)\, ,
\end{equation}		
the trace being taken into the Hilbert space of the $(0,2)$ superconformal theory corresponding 
to the compactification. This dressed elliptic genus, which 
is holomorphic in $z$ but not in $\tau$, transforms as a weak Jacobi form of weight $0$ and index $r/2$, where $r$ is the rank of the holomorphic vector bundle $\mathcal V$. 

The same non-holomorphic dressed elliptic genus can be defined for $K3\times T^2$, which corresponds to the particular case 
in which the torus fibration is trivial, hence the dressed elliptic genus factorizes into 
the usual elliptic genus of $K3$ and the partition function of the signature $(2,2)$ Narain lattice of the two-torus. 

Before quoting the result, let us summarize the relevant geometrical data characterizing the $\mathcal{N}=2$ compactifications of interest:
\begin{itemize}
\item A rank $r$ holomorphic vector bundle $\mathcal V$ over the (wrapped) $K3$ base $\mathcal{S}$, with $c_1 (\mathcal{V}) = 0$, whose pullback provides the 
gauge bundle of the compactification on $\mathcal{M}$. The structure group of $\mathcal V$ is embedded into the first $E_8$ factor, the second one being left unbroken.
\item A rational Narain lattice $\Gamma_{2,2}(T,U)$, since the two-torus moduli $T,U$ are quantized 
as a result of the presence of three-form flux, i.e. $T, U \in \mathbbm{Q}[\sqrt{D}]$ where $D$ is the discriminant of some positive-definite even quadratic form. This defines a $c=2$ toroidal rational conformal field theory~\cite{2004CMaPh.246..181G,Hosono:2002yb},
\item A pair of anti-self-dual two-forms $\omega_1$ and $\omega_2$ in $H^{2}(\mathcal{S},\mathbbm{Z})\cap 
\Lambda^{1,1} T_\mathcal{S}^\star$ characterizing the two-torus principal bundle.
\end{itemize}
This data is constrained by the Bianchi identity:
\begin{equation}
\label{eq:bianchi}
 \text{ch}_2(\mathcal{V})-\frac{T_2}{U_2}\,\omega\wedge\star_{\scriptscriptstyle{\mathcal{S}}}\bar\omega=\text{ch}_2(\mathcal T_\mathcal{S})\, ,
\end{equation}
which, upon integration over the base $\mathcal{S}$, gives the tadpole condition:
\begin{equation}
\label{eq:tadpole}
 n+\frac{T_2}{U_2}\int_\mathcal{S}\omega\wedge\star_{\scriptscriptstyle{\mathcal{S}}}\bar\omega=24\, .
\end{equation}
where:
\begin{equation}
n=-\int_\mathcal{S}\text{ch}_2(\mathcal{V})
\end{equation}
is the instanton number of the bundle $\mathcal{V}$, which is any integer between 0 and 24.

We also define a two-dimensional vector of two-forms $p_\omega$, built by embedding $(\omega_1,\omega_2)$ into 
the lattice of the two-torus fiber, and given in complex notation as:
\begin{equation}
\label{eq:def_pomega}
p_\omega:=\sqrt{\frac{T_2}{U_2}}\, (\omega_1+U\omega_2)\, .
\end{equation}
This vector belongs to a formal extension, over $H^2(\mathcal{S},\mathbbm{Z})$, 
of the winding sub-lattice of the $\Gamma_{2,2} (T,U)$ toroidal lattice.

The new supersymmetric index~(\ref{eq:susy_index_def}) was then computed in terms of this data in~\cite{Israel:2015aea,Israel:2016xfu}. First, 
the dressed elliptic genus~(\ref{eq:def_dressed}) can  be written as a sum of three terms in the following way:
\begin{align}
\label{eq:geometric_formula}
Z_\textsc{fy}=&\ \frac{1}{\eta(\tau)^{2}\bar{\eta}(\bar\tau)^{2}}
\sum_{\mu\in\Gamma_{\textsc{l}}^{\star}/\Gamma_{\textsc{l}}}
\sum_{\substack{p_\textsc{l}\in\Gamma_{\textsc{l}}+\mu\\p_\textsc{r}
\in\Gamma_{\textsc{r}}+\varphi(\mu)}}q^{\frac{1}{2}\langle 
p_\textsc{l},p_\textsc{l}\rangle_{\scriptscriptstyle
\Gamma_{\textsc{l}}}}
\bar{q}^{\frac{1}{2}\langle p_\textsc{r},p_\textsc{r}\rangle_{\scriptscriptstyle\Gamma_{\textsc{r}}}}\times\nonumber\\
&\times\left\{\frac{n}{24}
\left(\frac{\theta(\tau,z)}{\eta(\tau)}\right)^{r-2}Z_{\text{ell}}^{\scriptscriptstyle K3}(\tau,z)+\frac{\theta(\tau,z)^r}{12\, \eta(\tau)^{r+4}}\, (n-24)\hat E_2(\tau)\right.\nonumber\\
&\left.-\frac{\theta(\tau,z)^r}{2\, \eta(\tau)^{r+4}}\left(\int_{\mathcal{S}}\langle p_\textsc{l},p_\omega
\rangle_{\scriptscriptstyle\Gamma_{\textsc{l}}}^2-\frac{n-24}{2\pi\tau_2}\right)\right\}\, .
\end{align}
The definition of the various functions entering the above expression are summarized in \cref{app:hypergeom}. The left and right momenta $p_\textsc{l}$ and $p_\textsc{r}$ belong to the even lattices $\Gamma_\textsc{l}$ and $\Gamma_\textsc{r}$\footnote{These 
lattices are defined as 
$\Gamma_\textsc{l} = \Gamma_{2,2} (T,U) \cap \mathbbm{R}^{2,0}$ and $\Gamma_\textsc{l} = \Gamma_{2,2} (T,U) \cap \mathbbm{R}^{0,2}$ and 
are both of rank two because the corresponding $c=2$ CFT is rational.} 
shifted by $\mu$ and $\varphi(\mu)$ respectively, where $\mu$ is an element of the discriminant group 
$\Gamma_\textsc{l}^\star/\Gamma_\textsc{l}$ and 
$\varphi:\Gamma_\textsc{l}^\star/\Gamma_\textsc{l}\rightarrow\Gamma_\textsc{r}^\star/\Gamma_\textsc{r}$ is an isometry \cite{Hosono:2002yb}. In the above expression, $\langle\cdot,\cdot\rangle_{\scriptscriptstyle\Gamma}$ denotes the scalar product on the even lattice $\Gamma$.
We define then 
\begin{equation}
 f(p_\textsc{l},\omega):=\int_{\mathcal{S}}\langle p_\textsc{l},p_\omega\rangle_{\scriptscriptstyle\Gamma_{\textsc{l}}}^2-\frac{n-24}{2\pi\tau_2}= \int_{\mathcal{S}}\left(\langle p_\textsc{l},p_\omega\rangle_{\scriptscriptstyle\Gamma_{\textsc{l}}}^2-\frac{1}{4\pi\tau_2}\langle p_\omega,p_\omega\rangle_{\scriptscriptstyle\Gamma_{\textsc{l}}}\right)\, .
\label{eq:f_def}
\end{equation}
where we have used the tadpole condition \eqref{eq:tadpole}. 

Taking into account the remaining free fermions and performing the left GSO projection, one obtains then for the new supersymmetric index:
\begin{align}
 Z_\textsc{new}(\tau,\bar\tau)=&\ \frac{E_{4}(\tau)}{\eta(\tau)^{12}}
\sum_{\mu\in\Gamma_{\textsc{l}}^{\star}/\Gamma_{\textsc{l}}}
\sum_{\substack{p_\textsc{l}\in\Gamma_{\textsc{l}}+\mu\\p_\textsc{r}
\in\Gamma_{\textsc{r}}+\varphi(\mu)}}
q^{\frac{1}{2}\langle p_\textsc{l},p_\textsc{l}\rangle_{\scriptscriptstyle\Gamma_{\textsc{l}}}}
\bar{q}^{\frac{1}{2}\langle p_\textsc{r},p_\textsc{r}\rangle_{\scriptscriptstyle\Gamma_{\textsc{r}}}}\times\nonumber\\
 &\times\frac{1}{2}\sum_{\gamma,\delta=0}^{1}q^{\gamma^{2}}\left.\left\{\frac{n}{24}
\left(\frac{\theta(\tau,z)}{\eta(\tau)}\right)^{6}Z_{\text{ell}}^{\scriptscriptstyle K3}(\tau,z)+\right.\right. \\
 &\left.\left.+\frac{\theta(\tau,z)^8}{12\, \eta(\tau)^{12}}\, (n-24)\hat E_2(\tau)-
\frac{\theta(\tau,z)^8}{2\, \eta(\tau)^{12}}f(p_\textsc{l},\omega)\right\}\right|_{z=\frac{\gamma\tau+\delta}{2}}\, . 
\end{align}
Notice that the modular behaviour of the third term with a momentum insertion is ensured, since by construction the sum of 
the three terms is well-behaved and the first two terms are also by themselves weak 
almost holomorphic modular forms of weight $-2$.

Finally, in terms of standard weak almost holomorphic modular forms, the result takes a relatively simple form:
\begin{align}
\label{eq:newSusyIndex}
 Z_\textsc{new}(\tau,\bar\tau)=& \sum_{\mu\in\Gamma_{\textsc{l}}^{\star}/\Gamma_{\textsc{l}}}\sum_{\substack{p_\textsc{l}\in\Gamma_{\textsc{l}}+\mu\\p_\textsc{r}\in\Gamma_{\textsc{r}}+\varphi(\mu)}}
q^{\frac{1}{2}\langle p_\textsc{l},p_\textsc{l}\rangle_{\scriptscriptstyle\Gamma_{\textsc{l}}}}
\bar{q}^{\frac{1}{2}\langle p_\textsc{r},p_\textsc{r}\rangle_{\scriptscriptstyle\Gamma_{\textsc{r}}}}\ \times \nonumber\\
&\times \ \left(-\frac{n}{12}\frac{E_4 E_6}{\Delta}+\frac{n-24}{12}\frac{E_4^2 \hat{E}_2}{\Delta}
-\frac{f(p_\textsc{l},\omega)}{2}\frac{E_4^2}{\Delta}\right)\, , 
\end{align}
which will allow us to use the techniques developed in~\cite{Angelantonj:2011br,Angelantonj:2012gw,Angelantonj:2015rxa}, 
and reviewed briefly in the next section, to perform the integration over the fundamental domain of 
the worldsheet torus modular group leading to the various threshold corrections. 

The formula~(\ref{eq:geometric_formula}) that we used as a starting point was derived in~\cite{Israel:2016xfu} from a geometrical 
definition of the dressed elliptic genus, that coincides with the result obtained directly from a 
gauged linear sigma model using supersymmetric localization as we have proven there. We expect that this formula holds 
in full generality for all $\mathcal{N}=2$ compactifications with torsion of the class discussed in this work, even for those 
without an obvious GLSM realization. 

This result contains as a special case the standard $K3\times T^2$ compactifications, corresponding to the limiting case 
where the gauge instanton number $n$  equals  $24$ and where the momentum insertion $f(p_\textsc{l},\omega)$
 vanishes. 


\section{Niebur-Poincar\'e Series}
\label{sec:Niebur}

Integrals of the type
\begin{equation}
\int_\mathcal{F} \d \nu \, \varPhi (\tau ) \, \varLambda_{2,2} (T,U;\tau)  \label{modint}
\end{equation}
are quite common in string theory, since they compute the one-loop correction to couplings in the low-energy effective action. Here $d\nu = \d\tau_1 \d \tau_2 \, \tau_2^{-2}$ is the $SL (2, \mathbbm{Z})$ invariant measure, while $\mathcal{F} = \mathbbm{H} / SL (2,\mathbbm{Z}) $ is the fundamental domain of the modular group, $\mathbbm{H}$ being the Poincar\'e upper complex plane. $\varLambda_{2,2} (T,U;\tau) $ is the partition function associated to the $(2,2)$ dimensional Narain lattice, depending on the K\"ahler and complex structure moduli of the compactification torus as well as on the Teichm\"uller parameter $\tau$ of the worldsheet torus, while $\varPhi (\tau)$ is a, a priori, generic function invariant under the action of the modular group, whose explicit expression depends on the kind of coupling we are interested in. For those of interest in this paper, the automorphic function is weak quasi holomorphic modular function, in the sense that it has zero weight, it is holomorphic in the $\tau$ variable, aside from possible explicit $\tau_2$ dependence via the Eisenstein series $\hat E_2$, and has a simple pole at the cusp $\tau = i\infty$. Holomorphy is a consequence of the fact that the couplings we are interested in receive contributions only from BPS states. 

While the traditional way of computing the integral \eqref{modint} relies on the $SL (2,\mathbbm{Z})$ orbit decomposition of the Narain partition function, in \cite{Angelantonj:2011br, Angelantonj:2012gw, Angelantonj:2015rxa} a new method has been proposed whereby the fundamental domain is unfolded against the automorphic function $\varPhi$ itself. This way of proceeding has the clear advantage of keeping manifest the perturbative T-duality symmetries at all steps, and expresses the final result as a sum over BPS states. Moreover, singularities associated to states becoming massless at special points in the Narain moduli space are easily revealed in this representation.

In order to implement this strategy, it is essential that $\varPhi$ be represented as an absolutely convergent Poincar\'e series, so that the unfolding of the fundamental domain is justified. This is actually the case, since any weak quasi-holomorphic modular form can be uniquely decomposed in terms of so-called Niebur-Poincar\'e series $\mathcal{F} (s,\kappa , w)$, where $w$ is the modular weight, $\kappa$ determines the growth of the function at the cusp, while $s$ is a generic complex parameter. The Poincar\'e series representation of $\mathcal{F} (s,\kappa , w)$ is 
\begin{equation}
\mathcal{F} (s,\kappa , w) = \frac{1}{2} \sum_{(c,d)=1} (c\tau + d)^{-w}\, \mathcal{M}_{s,w} \left( - \frac{\kappa \tau_2}{|c\tau + d|^2}\right) \, \exp \left\{ - 2 i \pi \kappa \left( \frac{a}{c} - \frac{c \tau_1 + d}{c |c\tau + d|^2}\right)\right\}\,,
\end{equation}
in terms of the Whittaker $M$-function, $\mathcal{M}_{s,w} (y) = |4\pi y|^{-w/2} \, M_{\frac{w}{2}\textrm{sgn} (y) , s -\frac{1}{2}} (4 \pi |y|)$.

We refer the interested reader to \cite{Angelantonj:2011br, Angelantonj:2012gw, Angelantonj:2015rxa} for a general discussion of Niebur-Poincar\'e series. In the following we shall only remind that for negative weight, the choice $s=1-\frac{w}{2} +n$ is rather special, since the Niebur-Poincar\'e series are quasi holomorphic and absolutely convergent. As a result,
\begin{equation}
\varPhi (\tau ) = \sum_{n,\ell} d_\ell (n)\,   \mathcal{F} (1-\frac{w}{2} +n , \ell , w) \,,
\end{equation}
where the coefficients $d_\ell (n)$ are uniquely determined by matching the principal parts of the $q$-Laurent expansion of the two sides of the equation. In \cref{eq:decomp_Niebur1,eq:decomp_Niebur2,eq:decomp_Niebur3} we list the decomposition of interest for us, while we refer to \cite{Angelantonj:2012gw, Angelantonj:2015rxa} for more general cases. 

Since any weak quasi holomorphic modular form can be decomposed in terms of Niebur-Poincar\'e series, for the purpose of computing modular integrals it suffices to consider the basic integral 
\begin{equation}
I (s ) = \textrm{R.N.} \int_\mathcal{F} \d \nu \, \mathcal{F} (s,1,0) \, \varLambda_{2,2} (T,U; \tau )\,.
\end{equation}
Here we have selected  $\kappa =1$, the only case of interest in string theory. The symbol $\textrm{R.N.}$ (that we shall omit in the following, assuming that all integrals are properly renormalised) implies that the integral has been properly {\em renormalised} in order to cope with the infrared (logarithmic) divergences ascribed to massless states running in the loop. Our modular invariant prescription amounts at cutting-off the fundamental domain at large $\tau_2 > \mathcal{T}$, thus removing the singular behaviour of light states in the $\mathcal{T} \to \infty $ limit \cite{Angelantonj:2011br, Angelantonj:2012gw, Angelantonj:2015rxa}. 

Upon unfolding the fundamental domain against $\mathcal{F} (s,1,0)$ one gets \cite{Angelantonj:2012gw}
\begin{equation}
I (s) = \sum_{\textrm{BPS}} \int_0^\infty \frac{\d\tau_2}{\tau_2} \, \mathcal{M}_{s,0} (-\tau_2 ) \,e^{-\pi \tau_2 (p_\textrm{L}^2 +p_\textrm{R}^2 )/2} \label{modintunf}
\end{equation}
where the sum is restricted only to the BPS states satisfying $p_\textsc{l}^2 - p_\textsc{r}^2 =2$. The integral can be straightforwardly evaluated to yield \cite{Angelantonj:2012gw}
\begin{equation}
\label{eq:modular_integral}
\begin{split}
I_\alpha (s,w) &= \int_{\mathcal{F}}\, \text{d}\nu\,\, \tau_2^\alpha \sum_{p_\textsc{l},p_\textsc{r}}q^{\frac{1}{2}p_\textsc{l}^2}
\, \bar{q}^{\frac{1}{2}p_\textsc{r}^2}\, \mathcal{F}(s,\kappa,w)
\\
&=\ \sum_{p_\textsc{l},p_\textsc{r}}\delta\left(p_\textsc{l}^2-p_\textsc{r}^2-2\kappa\right)\left(4\pi\kappa\right)^{1-\alpha}
\left(\frac{p_\textsc{l}^2}{2\kappa}\right)^{-\frac{|w|}{2}-\alpha-s+1}\Gamma\left(\alpha+\frac{|w|}{2}+s-1\right)
\\
&\quad \times {}_2 F _1\left(\alpha+\frac{|w|}{2}+s-1,s-\frac{|w|}{2};2s;\frac{2\kappa}{p_\textsc{l}^2}\right)\, ,
\end{split}
\end{equation}
with $p_\textsc{l}^2:=\langle p_\textsc{l},p_\textsc{l}\rangle_{\scriptscriptstyle\Gamma_{\textsc{l}}}$ and $p_\textsc{r}^2:=\langle p_\textsc{r},p_\textsc{r}\rangle_{\scriptscriptstyle\Gamma_{\textsc{r}}}$, 
and
where we have allowed for a non-trivial weight of the Niebur-Poincar\'e series to compensate for Wilson lines and/or for momentum insertions in the Narain partition function \cite{Angelantonj:2012gw}. As we shall see in the next section, this representation of the modular integral clearly spells out possible IR divergences ascribed to new states becoming massless at points of symmetry enhancement. 

The integral \eqref{modintunf} can actually be given an alternative representation whenever the BPS constraint is solved before the $\tau_2$ integral is evaluated. The resulting representation defines a Fourier series expansion in the $T_1$ variable, which is only valid in special regions of moduli space (corresponding to large volume) \cite{Angelantonj:2015rxa}. For the case of momentum insertions we need to slightly generalise the construction of \cite{Angelantonj:2015rxa}, and we shall present the new results in section \ref{sec:Fourierseries}.


\section{Threshold corrections}

We are now ready to compute the one-loop threshold corrections to the gauge and gravitational coupling 
for $\mathcal{N}=2$ heterotic compactifications with torsion, starting from~(\ref{eq:newSusyIndex}) and using the 
techniques that were summarized in section~\ref{sec:Niebur}. Note that the actual models we are considering only exist at special points of the Narain moduli space compatible with the three-form flux. Nevertheless, we shall try to keep the moduli arbitrary and treat them as 
continuous variables, so that the expressions can be conveniently adapted to any special realization.\footnote{In particular, when $\omega_1$ and $\omega_2$ in~(\ref{eq:def_pomega}) are proportional to each other, only one complex torus modulus is stabilized by 
the flux and the other one remains.} 

\subsection{Gravitational threshold corrections}

In order to compute the threshold correction to the gravitational coupling, one has to compute the following modular integral:
\begin{equation}
 \Lambda_\text{grav}=\beta_\text{grav}\log\frac{M_s^2}{\mu^2}+\Delta_\text{grav}=\frac{1}{24}\int_{\mathcal{F}}\text{d}\nu\,\left\{\tau_2 \hat{E}_2(\tau)\, Z_\textsc{new}(\tau,\bar\tau)\right\}\, .
\end{equation}
Using \cref{eq:newSusyIndex}, one thus has to compute:
\begin{align}
 24\Lambda_\text{grav}=&\ \sum_{\mu,p_\textsc{l},p_\textsc{r}}\int_{\mathcal{F}}\text{d}\nu\,\, \tau_2\, 
q^{\frac{1}{2}\langle p_\textsc{l},p_\textsc{l}\rangle_{\scriptscriptstyle\Gamma_{\textsc{l}}}}
\bar{q}^{\frac{1}{2}\langle p_\textsc{r},p_\textsc{r}\rangle_{\scriptscriptstyle\Gamma_{\textsc{r}}}}\times\nonumber\\
&\times\left\{-\frac{n}{12}\frac{\hat{E}_2 E_4 E_6}{\Delta}+\frac{n-24}{12}\frac{\hat{E}_2^2 E_4^2}{\Delta}-\frac{f(p_\textsc{l},\omega)}{2}\frac{\hat{E}_2 E_4^2}{\Delta}\right\}\, .
\end{align}
where here and in the following, the momentum sum $\sum_{\mu,p_\textsc{l},p_\textsc{r}}$ is a compact notation for $\sum_{\mu\in\Gamma_{\textsc{l}}^{\star}/\Gamma_{\textsc{l}}}\sum_{p_\textsc{l}\in\Gamma_{\textsc{l}}+\mu}
\sum_{p_\textsc{r}\in\Gamma_{\textsc{r}}+\varphi(\mu)}$.

Following~\cite{Angelantonj:2012gw} we rewrite the weak almost holomorphic modular forms entering in the integrands above in terms of Niebur-Poincar\'e series $\mathcal{F}(s,\kappa,w)$.
One has the following decompositions:
\begin{equation}
\label{eq:decomp_Niebur1}
\begin{split}
\frac{\hat E_2 E_4 E_6}{\varDelta} &= \cF (2,1,0) - 5 \cF(1,1,0)-144\,,
\\
\frac{{\hat E_2}^2 E_4^2}{\varDelta} &= \tfrac{1}{5} \cF (3,1,0) - 4 \cF (2,1,0) + 13 \cF (1,1,0) + 144\,,
\\
\frac{\hat E_2 E_4^2}{\varDelta} &= \tfrac{1}{40} \cF (3,1,-2) - \tfrac{1}{3} \cF (2,1,-2)\,.
\end{split}
\end{equation}
Regularizing the IR divergence and performing the modular integral by unfolding the integration domain against the Niebur-Poincar\'e series, one obtains 
\begin{align}
\Lambda_\text{grav}=&\ \sum_{\text{BPS}}\Bigg\{-\frac{m(p_\textsc{l})}{48}  \left(\frac{3 \ _2 F_1\left(2,4,6,t^{-1}\right)}{20 t^4}-\frac{2 \ _2 F_1\left(1,3,4,t^{-1}\right)}{3 t^3}\right)\nonumber\\
&-\frac{n}{12\times 24} \left(\frac{\ _2 F_1\left(2,2,4,t^{-1}\right)}{t^2}-\frac{5 \ _2 F_1\left(1,1,2,t^{-1}\right)}{t}\right)\nonumber\\
&+\frac{n-24}{24} \left(\frac{\ _2 F_1\left(2,3,6,t^{-1}\right)}{20 t^3}-\frac{\ _2 F_1\left(1,2,4,t^{-1}\right)}{3 t^2}\right)\nonumber\\
&+\frac{n-24}{12\times 24} \left(\frac{2 \ _2 F_1\left(3,3,6,t^{-1}\right)}{5 t^3}-\frac{4 \ _2 F_1\left(2,2,4,t^{-1}\right)}{t^2}+\frac{13 \ _2 F_1\left(1,1,2,t^{-1}\right)}{t}\right)\Bigg\}\nonumber\\
&+(n-12)\mathcal{I}_{\textsc{dkl}}\,,
\end{align}
where $t:=p_{\textsc{l}}^2/2$,  $m(p_\textsc{l},\omega):=\int_{\mathcal{S}}\langle p_{\textsc{l}},p_\omega\rangle_{\Gamma_{\textsc{l}}}^2$ and 
\begin{equation}
\label{eq:torus_integral}
\mathcal{I}_{\textsc{dkl}}:=\int_{\mathcal{F}}\, \text{d}\nu\,\, \tau_2\sum_{\mu,p_\textsc{l},p_\textsc{r}}q^{\frac{1}{2}p_\textsc{l}^2}
\, \bar{q}^{\frac{1}{2}p_\textsc{r}^2}=-\text{log}\left(4\pi e^{-\gamma}T_2 U_2\left|\eta(T)\eta(U)\right|^4\right)
\end{equation}
is the Dixon-Kaplunovsky-Louis integral, where $\gamma$ is the Euler-Mascheroni constant.
As already explained, in the above expression, $\sum_{\text{BPS}}$ is a shorthand for $\sum_{p_\textsc{l},p_\textsc{r}}\delta(p_\textsc{l}^2-p_\textsc{r}^2-2)$, 
in other words the sum over perturbative half-BPS states.

Fortunately, this complicated expression simplifies considerably in the cases of interest here, and one ends up with standard polynomial and logarithmic functions of the last argument, cf. \cref{app:hypergeom}. One ends up with the following simple expression for the gravitational threshold corrections:
\begin{align}
\label{eq:resultThreshGrav}
\Lambda_\text{grav}=&\ \sum_{\text{BPS}}\Bigg\{1+\frac{n-24}{24}\,\frac{3}{2t}+\left(t-\frac{11}{12}\right)\log\left(\frac{t-1}{t}\right)+\nonumber\\
&+\frac{m(p_\textsc{l},\omega)}{24}\left[6-\frac{3}{4t^2}-\frac{5}{2t}+6\left(t-\frac{11}{12}\right)\log\left(\frac{t-1}{t}\right)\right]\Bigg\}\nonumber\\
&+(n-12)\, \mathcal{I}_{\textsc{dkl}}\,.
\end{align}
This expression is clearly independent of the choice of chamber in the Narain moduli space. 

Setting $n=24$ and $m(p_\textsc{l},\omega)=0$ to make the torus fibration trivial, 
one obtains the result for $K3\times T^2$ compactifications:
\begin{equation}
\label{eq:grav_beta}
\Lambda_\text{grav}=\sum_{\text{BPS}}\left\{1+\left(t-\frac{11}{12}\right)\log\left(\frac{t-1}{t}\right)\right\}+12\, \mathcal{I}_{\textsc{dkl}}\, .
\end{equation}
Note that these expressions are potentially divergent if $t=1$, {\em i.e.} at point of symmetry enhancement where $p_\text{L}^2 =2$. The presence or not of these divergences clearly depends of the actual values of the K\"ahler and complex structure moduli.

Finally from~\cref{eq:resultThreshGrav} we can extract the value of gravitational $\beta$-function, which is the coefficient of 
the trace anomaly~\cite{Antoniadis:1992sa}:
\begin{equation}
\beta_\text{grav}=n-12\,.
\end{equation}
This coefficient is related to the relative number of hypermultiplets and vector multiplets. 
Comparing eq.~(\ref{eq:grav_beta}) with known results from $K3\times T^2$~\cite{Kiritsis:1997hj}, we get the 
normalisation:
\begin{equation}
\beta_\text{grav}=\frac{24+n_\textsc{h}-n_\textsc{v}}{22}\,.
\end{equation}
Hence $n_\textsc{h}-n_\textsc{v}$, $i.e.$ the difference between the number of massless hypermultiplets and vector 
multiplets (including $S$, containing the dilaton, and the graviphoton), 
depends on the instanton number $n$ of the vector bundle $\mathcal{V}$, hence indirectly on the 
data of the principal two-torus bundle through the integrated Bianchi identity~(\ref{eq:tadpole}).


\subsection{Gauge threshold corrections}

The expression~(\ref{eq:newSusyIndex}) for the new supersymmetric index is independent of the rank of the gauge bundle. In order to compute explicitely the correction to the gauge couplings one has to choose a particular sub-class of bundles; we will consider below the case of a bundle of structure group $SU(2)$, embedded into one of the two $E_8$ factors of the gauge group, with 
arbitary instanton number $0\leqslant n \leqslant 24$. It will allow 
to compare easily with classical results for $K3 \times T^2$ with the standard embedding of the spin connection into the gauge connection, and vanishing Wilson lines, $i.e.$ models with a rank one bundle and $n=24$.  

\subsubsection{Corrections to the \texorpdfstring{$E_8$}{E8} coupling}

Let us start with the one-loop correction to the gauge coupling corresponding to 
the unbroken $E_8$ factor of the spacetime gauge group. The threshold correction is given by:
\begin{equation}
\Lambda_{E_8}=\beta_{E_8}\log\frac{M_s^2}{\mu^2}+\Delta_{E_8}=\int_{\mathcal{F}}\, \text{d}\nu\, Z_{E_8}(\tau,\bar\tau)\, ,
\end{equation}
where $Z_{E_8}(\tau,\bar\tau)$ corresponds to the new supersymmetric index with an extra insertion of $\left(Q_{E_8}^2-\frac{1}{8\pi\tau_2}\right)$ in the trace:
\begin{equation}
Z_{E_8}(\tau,\bar\tau)=\frac{\tau_2}{\eta(\tau)^2}\,\text{Tr}_\textsc{r}\left\{\left(Q_{E_8}^2-\frac{1}{8\pi\tau_2}\right)\bar{J}_{\, 0}
(-1)^{F_\textsc{r}} q^{L_0-c/24}\bar{q}^{\bar{L}_0-\bar{c}/24}\right\}
\end{equation}

Let us define $\tilde D_w:=(-4w)^{-1}D_w$, where $D_w$ is the modular covariant derivative as defined in \cref{app:hypergeom}. The insertion $\left(Q_{E_8}^2-\frac{1}{8\pi\tau_2}\right)$ corresponds then to acting in $Z_\textsc{new}$ on the character of the affine $E_8$ algebra, namely $E_4(\tau)$ with the operator $\tilde D_4$. Using the fact that:
\begin{equation}
D_4E_4=\frac{2}{3}\left(E_6-\hat E_2 E_4\right)\, ,
\end{equation}
One obtains:
\begin{equation}
Z_{E_8}=\frac{\hat E_2 E_4-E_6}{24\Delta}\,\tau_2\sum_{\mu,p_\textsc{l},p_\textsc{r}}q^{\frac{1}{2}p_\textsc{l}^2}
\, \bar{q}^{\frac{1}{2}p_\textsc{r}^2}\left\{-\frac{n}{12}E_6+\frac{n-24}{12}\hat E_2 E_4-\frac{f(p_\textsc{l},\omega)}{2}E_4\right\}\, ,
\end{equation}
i.e.:
\begin{equation}
\begin{split}
Z_{E_8}=&\ \frac{1}{24\Delta}\,\tau_2\sum_{\mu,p_\textsc{l},p_\textsc{r}}q^{\frac{1}{2}p_\textsc{l}^2}
\, \bar{q}^{\frac{1}{2}p_\textsc{r}^2}\times\\
&\left\{-\frac{n-12}{6}\hat E_2E_4E_6+\frac{n}{12}E_6^2+\frac{n-24}{12}\hat E_2^2E_4^2-\frac{f(p_\textsc{l},\omega)}{2}(\hat E_2E_4^2-E_4E_6)\right\}\, .
\end{split}
\end{equation}
In addition to \cref{eq:decomp_Niebur1}, one has the following decompositions into Niebur-Poincar\'e series:
\begin{equation}
\label{eq:decomp_Niebur2}
\begin{split}
\frac{E_6^2}{\Delta}&=\mathcal{F}(1,1,0)-1008\, ,\\
\frac{E_4E_6}{\Delta}&=\tfrac{1}{6}\mathcal{F}(2,1,-2)\, .
\end{split}
\end{equation}
One then performs the modular integral to get:
\begin{align}
\Lambda_{E_8}=&\ \sum_{\text{BPS}}\Bigg\{-\frac{m(p_\textsc{l})}{48} \left(\frac{3 \ _2 F_1\left(2,4,6,t^{-1}\right)}{20 t^4}-\frac{\ _2 F_1\left(1,3,4,t^{-1}\right)}{t^3}\right)\nonumber\\
&-\frac{n-12}{6\times 24} \left(\frac{\ _2 F_1\left(2,2,4,t^{-1}\right)}{t^2}-\frac{5 \ _2 F_1\left(1,1,2,t^{-1}\right)}{t}\right)\nonumber\\
&+\frac{n-24}{24} \left(\frac{\ _2 F_1\left(2,3,6,t^{-1}\right)}{20 t^3}-\frac{\ _2 F_1\left(1,2,4,t^{-1}\right)}{2 t^2}\right)+\frac{n \ _2 F_1\left(1,1,2,t^{-1}\right)}{12 t}\nonumber\\
&+\frac{n-24}{12\times 24}  \left(\frac{2 \ _2 F_1\left(3,3,6,t^{-1}\right)}{5 t^3}-\frac{4 \ _2 F_1\left(2,2,4,t^{-1}\right)}{t^2}+\frac{13 \ _2 F_1\left(1,1,2,t^{-1}\right)}{t}\right)\Bigg\}\nonumber\\
&-2(n+12)\, \mathcal{I}_{\textsc{dkl}}\, .
\end{align}
Once again, for such integer values of the arguments, the hypergeometric functions simplify dramatically, cf. \cref{eq:Hypergeom_Simple}, and one ends up with the following simple expression:
\begin{align}
\Lambda_{E_8}=&\ \sum_{\text{BPS}}\Bigg\{1+\frac{n-24}{12t}+(t-1)\log\left(\frac{t-1}{t}\right)+\nonumber\\
&+\frac{m(p_\textsc{l},\omega)}{4}\left[1-\frac{1}{6t^2}-\frac{1}{2t}+(t-1)\log\left(\frac{t-1}{t}\right)\right]\Bigg\}\nonumber\\
&-2(n+12)\, \mathcal{I}_{\textsc{dkl}}\, .
\end{align}
From this expression we can read off immediately the expression of the $\beta$-function:
\begin{equation}
\beta_{E_8}=-2(n+12)\,.
\end{equation}
Setting $n=24$ and $m(p_\textsc{l},\omega)=0$, one obtains:
\begin{equation}
\Lambda_{E_8}=\sum_\text{BPS}\left\{1+(t-1)\log\left(\frac{t-1}{t}\right)\right\}-72\, \mathcal{I}_{\textsc{dkl}}\,,
\end{equation}
which coincides with the already known result for $K3\times T^2$ \cite{Angelantonj:2012gw}.


\subsubsection{Corrections to the \texorpdfstring{$E_7$}{E7} coupling}

For definiteness, and as stated in the introduction of this section, we focus on the case in which the vector bundle 
over the compact manifold has an $SU(2)$ structure group, such that the unbroken gauge group in spacetime contains a $E_7$ factor. 

As before, computing the threshold correction corresponds to performing the modular integral of a descendant of the new supersymmetric index, i.e. with a $\left(Q_{E_7}^2-\frac{1}{8\pi\tau_2}\right)$ insertion in the trace:
\begin{equation}
\Lambda_{E_7}=\beta_{E_7}\log\frac{M_s^2}{\mu^2}+\Delta_{E_7}=\int_{\mathcal{F}}\, \text{d}\nu\, Z_{E_7}(\tau,\bar\tau)\, ,
\end{equation}
with:
\begin{equation}
Z_{E_7}(\tau,\bar\tau)=\frac{\tau_2}{\eta(\tau)^2}\,\text{Tr}_\textsc{r}\left\{\left(Q_{E_7}^2-\frac{1}{8\pi\tau_2}\right)\bar{J}_{\, 0}
(-1)^{F_\textsc{r}} q^{L_0-c/24}\bar{q}^{\bar{L}_0-\bar{c}/24}\right\}
\end{equation}
In functional picture, the extra operator insertion acts as $\tilde D_w$ on every $E_4(\tau)$ and $E_6(\tau)$ factor in the new supersymmetric index but not on the $E_4(\tau)$ corresponding to the unbroken $E_8$ factor of the gauge group, which was treated in the previous section. One has the following identities, due to Ramanujan:
\begin{subequations}
\begin{align}
D_4E_4&=\frac{2}{3}\left(E_6-\hat E_2 E_4\right)\, ,\\
D_6E_6&=E_4^2-\hat E_2 E_6\, .
\end{align}
\end{subequations}
One thus obtains:
\begin{align}
Z_{E_7}=&\ \frac{\tau_2}{24\Delta}\sum_{\mu,p_\textsc{l},p_\textsc{r}}q^{\frac{1}{2}p_\textsc{l}^2}
\, \bar{q}^{\frac{1}{2}p_\textsc{r}^2}\times\\
&\left\{-\frac{n-12}{6}\hat E_2E_4E_6+\frac{n}{12}E_4^3+\frac{n-24}{12}\hat E_2^2E_4^2-\frac{f(p_\textsc{l},\omega)}{2}(\hat E_2E_4^2-E_4E_6)\right\}\, .
\end{align}
In addition to \cref{eq:decomp_Niebur1,eq:decomp_Niebur2} one has the following decomposition into Niebur-Poincar\'e series:
\begin{equation}
\label{eq:decomp_Niebur3}
\begin{split}
\frac{E_4^3}{\Delta}=\mathcal{F}(1,1,0)+720\, .
\end{split}
\end{equation}
It gives:
\begin{align}
\Lambda_{E_7}=&\ \sum_{\text{BPS}}\Bigg\{-\frac{m(p_\textsc{l})}{48} \left(\frac{3 \ _2 F_1\left(2,4,6,t^{-1}\right)}{20 t^4}-\frac{\ _2 F_1\left(1,3,4,t^{-1}\right)}{t^3}\right)\nonumber\\
&+\frac{n}{12\times 24} \frac{\ _2 F_1\left(1,1,2,t^{-1}\right)}{t}+\frac{n-24}{24} \left(\frac{\ _2 F_1\left(2,3,6,t^{-1}\right)}{20 t^3}-\frac{\ _2 F_1\left(1,2,4,t^{-1}\right)}{2 t^2}\right)\nonumber\\
&+\frac{n-24}{12\times 24}  \left(\frac{2\ _2 F_1\left(3,3,6,t^{-1}\right)}{5 t^3}-\frac{4 \ _2 F_1\left(2,2,4,t^{-1}\right)}{t^2}+\frac{13 \ _2 F_1\left(1,1,2,t^{-1}\right)}{t}\right)\Bigg\}\nonumber\\
&+4(n-6)\mathcal I_{\textsc{dkl}}\, .
\end{align}
Once again, for such integer values of the arguments, the hypergeometric functions simplify dramatically, cf. \cref{eq:Hypergeom_Simple}, and one ends up with the following simple expression:
\begin{align}
\Lambda_{E_7}=&\ \sum_{\text{BPS}}\Bigg\{1+\frac{n-24}{12t}+(t-1)\log\left(\frac{t-1}{t}\right)\nonumber\\
&+\frac{m(p_\textsc{l},\omega)}{4}\left[1-\frac{1}{6t^2}-\frac{1}{2t}+(t-1)\log\left(\frac{t-1}{t}\right)\right]\Bigg\}\nonumber\\
&+4(n-6)\, \mathcal{I}_{\textsc{dkl}}\, .
\end{align}
We can once again read directly the $\beta$-function:
\begin{equation}
\beta_{E_7}=4(n-6)\,.
\end{equation}
Setting $n=24$ and $m(p_\textsc{l},\omega)=0$, one obtains:
\begin{equation}
\Lambda_{E_7}=\sum_\text{BPS}\left\{1+(t-1)\log\left(\frac{t-1}{t}\right)\right\}+72\, \mathcal{I}_{\textsc{dkl}}\,,
\end{equation}
corresponding indeed to the already known result for $K3\times T^2$.


\subsubsection{Universality property of the gauge threshold corrections}
The presence of $\mathcal{N}=2$ supersymmetry in spacetime hints towards some universality properties of the thresholds, 
as in the $K3\times T^2$ case. The difference of the two gauge thresholds indeed turns out to be universal. Using the fact that:
\begin{equation}
E_4^3-E_6^2=1728\Delta\, ,
\end{equation}
one obtains for the difference of the two integrands:
\begin{equation}
Z_{E_8}-Z_{E_7}=-6n\,\tau_2\sum_{\mu,p_\textsc{l},p_\textsc{r}}q^{\frac{1}{2}p_\textsc{l}^2}
\, \bar{q}^{\frac{1}{2}p_\textsc{r}^2}\, 
\end{equation}
leading to an integer multiple of the Dixon-Kaplunovsky-Louis integral for the thresholds:
\begin{equation}
\Lambda_{E_8}-\Lambda_{E_7}=(\beta_{E_8}-\beta_{E_7})\, \mathcal{I}_{\textsc{dkl}}=-6n\, \mathcal{I}_{\textsc{dkl}}\, .
\end{equation}
Setting $n=24$, one recovers the well-known result:
\begin{equation}
\Lambda_{E_8}-\Lambda_{E_7}=-144\, \mathcal{I}_{\textsc{dkl}}\, .
\end{equation}


\section{Fourier series and worldsheet instanton corrections}\label{sec:Fourierseries}

The results obtained in the previous section encapsulate in a compact and $O(2,2;\mathbbm{Z})$ invariant way the threshold corrections 
to the gauge and gravitational couplings. It is useful to present the result in a different way, which allows one to isolate 
the contributions from worldsheet instantons, using a Fourier series expansion~\cite{Angelantonj:2015rxa}. 

The role of worldsheet instantons is particularly interesting to investigate in these $\mathcal{N}=2$ torsional 
compactifications, whose topology corresponds to the total space of the principal bundle $T^2\hookrightarrow\mathcal M\stackrel{\pi}{\to}\mathcal{S}$. The relevant instantons in this context are holomorphic maps from the worldsheet two-torus to the target-space $T^2$. 

In the present context neither the $K3$ base nor the $T^2$ fiber are cycles 
 of the total space $\mathcal{M}$ of the principal bundle; in particular the two-torus is only a torsion two-cycle. 
One may wonder therefore whether an infinite tower of instanton corrections appears in the result; as 
we will see below, it turns out to be the case. 

Starting from $\text{Spin}(32)/\mathbbm{Z}_2$ ten-dimensional heterotic strings, our results lead to interesting insights 
on non-perturbative corrections to Type I compactifications with Ramond-Ramond flux. Under heterotic/type I S-duality, 
the one-loop heterotic computations capture both perturbative and non-pertubative corrections on the type I side, in particular 
the contribution of Euclidean D1-brane worldsheets wrapping the two-torus~\cite{Kiritsis:1999ss,Camara:2008zk}. This is a quite interesting result, as D-instantons corrections in the presence of RR fluxes have not been investigated in detail to our knowledge. 

\subsection{The Fourier series expansion}
\label{sec:FourierSeries}

Let us now focus on an alternative representation in terms of a Fourier series expansion of the integral:
\begin{equation}
\mathcal{I}_g(s):=\int_\mathcal{F}\text{d}\nu\,\tau_2\sum_{\mu,p_\textsc{l},p_\textsc{r}}g(p_\textsc{l}\sqrt{\tau_2})\,q^{\frac{1}{2}p_\textsc{l}^2}\bar q^{\frac{1}{2}p_\textsc{r}^2}\,\mathcal{F}(s,1,w)\,,
\end{equation}
with some momentum insertion $g(p_\textsc{l}\sqrt{\tau_2})$, which in our case will correspond to $f(p_\textsc{l},\omega)$.

In order to obtain this alternative Fourier series representation, one first performs the $\tau_1$ integral which imposes the BPS constaint on the momenta, then solves the constraint and performs a suitable Poisson resummation.

Explicitly, one expands the Niebur-Poincar\'e series in terms of the Whittaker M-function, which is then itself expressed in terms of the confluent hypergeometric function ${}_1 F_1$,
\begin{equation}
\begin{split}
\cM_{s,w} (-t) &= (4\pi t )^{-w/ 2} M_{-w/2,s-1/2} (4\pi t )  
\\
&= (4\pi t )^{s-w/2} \, e^{-2\pi t} \, {}_1 F_1 (s+w/2;2s;4\pi t)\,.
\end{split}
\end{equation}
The hypergeometric function ${}_1 F_1$ satisfies:
\begin{equation}
{}_1 F_1 (a;2a+n;y) = \varGamma (a-\tfrac{1}{2}) \left( \frac{y}{4}\right)^{\frac{1}{2}-a}\, e^{y/2}\, \sum_{\ell =0}^n \frac{ (-n)_\ell \, (2a-1)_\ell}{(2a+n)_\ell\, \ell!} (a+\ell-\tfrac{1}{2})\, I_{a+\ell - \frac{1}{2} } (y/2)\,.\label{1F1Bessel}
\end{equation}
In these expressions $(x)_\ell = \varGamma (x + \ell ) /\varGamma (x) = x (x+1 )\ldots (x+l-1)$ is the Pochhammer symbol  or rising factorial. It satisfies, $(-x)_\ell = (-1)^\ell (x-\ell+1)_\ell$. 

This strategy can be applied first to compute the Fourier series expansion in absence of momentum insertion:

\begin{equation}
\label{eq:FourierIntegral}
\mathcal{I}(s):=\int_\cF \text{d}\nu \, \cF (s,1,0)\, \varLambda_{2,2} (T,U) \, ,
\end{equation}
where:
\begin{equation}
\varLambda_{2,2} (T,U):=\tau_2\sum_{\mu,p_\textsc{l},p_\textsc{r}}q^{\frac{1}{2}p_\textsc{l}^2}\bar q^{\frac{1}{2}p_\textsc{r}^2}
\end{equation}
is the modular invariant partition function of the signature $(2,2)$ Narain lattice. It is evaluated at some particular 
points in moduli space specified by the quantization condition $T,U\in\mathbbm{Q}[\sqrt{D}]$, although the computation below,
by itself, could be done for any $T$ and $U$ as nowhere we make use of these conditions. 

The Fourier expansion of \cref{eq:FourierIntegral} was computed in \cite{Angelantonj:2015rxa}. 
The result splits into zero, negative and positive frequency parts:
\begin{equation}
\mathcal{I}(s)=\mathcal{I}^{(-)}(s)+\mathcal{I}^{(0)}(s)+\mathcal{I}^{(+)}(s)\, ,
\end{equation}
with:
\begin{equation}
\label{eq:intWithoutInsertion}
\begin{split}
\mathcal{I}^{(0)}(s)=&\ 2^{4s-3}\sqrt{4\pi}\Gamma\left(s-\frac{1}{2}\right)\sum_{(n^1,n^2)=1}(T_2\tilde U_2)^s\left(T_2+\tilde U_2+|T_2-\tilde U_2|\right)^{1-2s}\, ,\\
\mathcal{I}^{(+)}(s)=&\, \frac{1}{2}\sum_{M>0}\sum_{\gamma\in\Gamma_\infty\backslash\Gamma_U}\frac{e^{2i\pi M(T_1-\tilde U_1)}}{M}\mathcal{M}_{s,0}\left(\frac{M}{2}\left(T_2+\tilde U_2-|T_2-\tilde U_2|\right)\right)\times\\
&\times\mathcal{W}_{s,0}\left(\frac{M}{2}\left(T_2+\tilde U_2+|T_2-\tilde U_2|\right)\right)\, ,
\end{split}
\end{equation}
the negative frequency part being obtained by complex conjugation.

Using the relations between the Whittaker functions $M_{k,m}$, $W_{k,m}$ 
and the modified Bessel functions of the first and second kind \cite{Angelantonj:2012gw}:
\begin{equation}
\begin{split}
\mathcal{M}_{s,0}(\pm y)&=2^{2s-1}\Gamma\left(s+\frac{1}{2}\right)\sqrt{4\pi|y|}\, I_{s-\frac{1}{2}}(2\pi|y|)\,,\\
\mathcal{W}_{s,0}(\pm y)&=2\sqrt{|y|}\, K_{s-\frac{1}{2}}(2\pi|y|)\,,
\end{split}
\end{equation}

and focusing on the fundamental chamber $T_2>\tilde U_2$ for definiteness, one can rewrite the positive frequency part in the following way:

\begin{equation}
\label{eq:posfreqpart}
\begin{split}
\mathcal{I}^{(+)}(s)=&\,2^{2s+1}\sqrt{\pi}\,\Gamma\left(s+\frac{1}{2}\right)\times\\
&\times\sum_{M>0}\sum_{\gamma\in\Gamma_\infty\backslash\Gamma}\sqrt{T_2\tilde U_2}\,e^{2i\pi M(T_1-\tilde U_1)}I_{s-\frac{1}{2}}(2\pi M\tilde U_2)K_{s-\frac{1}{2}}(2\pi MT_2)\,,
\end{split}
\end{equation}
where one recognizes the sum over comprime integers $n^1,n^2$ as a sum over cosets in the quotient of 
the modular group $\Gamma=SL_2(\mathbbm{Z})_U$ by the stabilizer of the cusp at infinity 
$\Gamma_\infty$. Notice the presence of a factor of 2 since the pairs $(n^1,n^2)$ and $(-n^1,-n^2)$ 
correspond to the same coset $\gamma$.

We now want to compute the Fourier series expansion of an integral of the same 
type but with the extra $f(p_\textsc{l},\omega)$ weight 2 momentum insertion, namely:
\begin{equation}
\label{eq:intWithInsertion}
\mathcal{I}_f(s):=\int_\cF \text{d}\nu \, \cF (s,1,-2)\, \tau_2\sum_{\mu,p_\textsc{l},p_\textsc{r}}q^{\frac{1}{2}p_\textsc{l}^2}\bar q^{\frac{1}{2}p_\textsc{r}^2}
f(p_\textsc{l},\omega)\,,
\end{equation}
with:
\begin{equation}
f(p_\textsc{l},\omega) = \int_{\mathcal S}\langle p_\textsc{l},p_\omega\rangle_{\Gamma_\textsc{l}}^2 -\frac{n-24}{2\pi\tau_2}\,.
\end{equation}
Upon unfolding the fundamental domain $\cF$ against the Niebur-Poincar\'e series\newline $\cF (s,1,-2)$ one gets:
\begin{align}
\label{eq:IntegralWhittaker}
\mathcal{I}_f(s)&= \int_0^\infty \frac{\text{d}\tau_2}{\tau_2^2} \int_{-1/2}^{1/2} \text{d}\tau_1 \, \cM_{s,-2 } (\tau_2 ) e^{-2\pi i \tau_1} \, \sum_{\mu,p_\textsc{l},p_\textsc{r}}q^{\frac{1}{2}p_\textsc{l}^2}\bar q^{\frac{1}{2}p_\textsc{r}^2}
f(p_\textsc{l},\omega)\nonumber
\\
& = \sum_\textsc{bps} \int_0^\infty \frac{\text{d}\tau_2}{\tau_2} \, \cM_{s,-2} (\tau_2 ) f(p_\textsc{l},\omega) \, e^{-\pi \tau_2 (|p_\textsc{l}|^2 + |p_\textsc{r}|^2 )/2}\,.
\end{align}
The $\tau_1$ integration variable acts as a Lagrange multiplier to restrict the lattice sum to the contributions $m_1 n^1 + m_2 n^2 =1$, where we have expanded the momenta in a complex basis:
\begin{subequations}
\begin{align}
p_L &= \frac{1}{\sqrt{U_2 T_2}} \left(m_2 - U m_1 + \bar{T} (n^1 + U n^2) \right)\\
p_R &= \frac{1}{\sqrt{U_2 T_2}} \left(m_2 - U m_1 + T (n^1 + U n^2) \right)
\end{align}
\end{subequations}
 
As explained above, first one has to solve the BPS constraint $m_1 n^1 + m_2 n^2 =1$. In general, for any pair of co-prime integers $(n^1 , n^2)$, B\'ezout's lemma ensures that one can find another pair of co-primes $(\tilde m_1 , \tilde m_2) $ such that $\tilde m_1 n^1 + \tilde m_2 n^2 =1$. The solutions of the BPS constraints are then of the form:
\begin{equation}
\begin{split}
m_1 &= \tilde m_1 + \tilde M n^2\,,\\
m_2 &= \tilde m_2 - \tilde M n^1\,,
\end{split}
\end{equation}
with $\tilde M\in \mathbbm{Z}$.
Upon inserting this expression into the integrand one notices that the complex structure $U$ and the charges defining $p_\omega$ always appear in the combination $\tilde U = \gamma \cdot U$ so that the sum over $(n^1 , n^2)$ reduces to a sum over images with respect to $SL(2;\mathbbm{Z})_U$. At this point one has to Poisson resum over the variable $\tilde M$ to obtain the desired Fourier series expansion in $T$. Notice that the momenta are at most linear in $\tilde M$ which imply that both the argument in the exponential and the $f(p_\textsc{l},\omega)$ insertion in \cref{eq:IntegralWhittaker} are polynomials of second degree in $\tilde M$. One gets the following schematic expression for this integral \cref{eq:intWithInsertion}:

\begin{equation}
\label{eq:intWithInsertion2}
\begin{split}
\mathcal{I}_f(s)=2\sum_{M\in\mathbbm{Z}}\sum_{\gamma\in\Gamma_\infty\backslash\Gamma}&\sqrt{T_2\tilde U_2}e^{2i\pi M(T_1-\tilde{U_1})}\times\\
&\times\int_0^\infty\frac{\text{d}\tau_2}{\tau_2^{3/2}}\,\mathcal{M}_{s,-2}(-\tau_2)\left(\frac{\alpha}{\tau_2^2}+\frac{\beta}{\tau_2}+\delta\right)\exp\left(-\frac{A}{\tau_2}-B\tau_2\right)\,.
\end{split}
\end{equation}
The precise expression of the various coefficients in the above schematic expression is determined in \cref{app:coeff}. We recall that
\begin{equation}
\begin{split}
\cM_{s,-2} (y ) &= 4 \pi y \, M_{1,s-1/2} (4\pi y)
\\
&= (4\pi y)^{s+1} \, e^{-2\pi y} \, {}_1 F_1 (s-1;2(s-1)+2 ; 4\pi y )\,,
\end{split}
\end{equation}
that, together with eq. \eqref{1F1Bessel} yields:
\begin{equation}
\label{eq:WhittakerBessel}
\cM_{s,-2} (y) = 2^{2s-3} \, \varGamma (s-\tfrac{3}{2}) \, (4\pi y )^{5/2} \, \sum_{\ell =0}^2 (-1)^\ell \frac{(3-\ell)_\ell \, (2s-3)_\ell}{(2s)_\ell \, \ell!} \, (s+\ell -\tfrac{3}{2})\, I_{s+\ell - \frac{3}{2}} (2\pi y)\,.
\end{equation}
Plugging this expression into the integral in \cref{eq:intWithInsertion2} yields:

\begin{align}
4^{2s} \,\pi^{5/2} \, \varGamma (s-\tfrac{3}{2}) \, \sum_{\ell =0}^2 (-1)^\ell &\frac{(3-\ell)_\ell \, (2s-3)_\ell}{(2s)_\ell \, \ell!} \, (s+\ell -\tfrac{3}{2})\times\nonumber\\
&\times \int_0^\infty \text{d}t\, \left( \frac{\alpha}{t} +\beta+\delta t\right)\, I_{s+\ell -\frac{3}{2}} (2 \pi t )\, e^{-A/t - B t} \label{result}
\,.
\end{align}

The relevant values of the coefficients $A,B,\alpha,\beta$ and $\delta$ computed in \cref{app:coeff} are the following:
\begin{subequations}
\begin{align}
A&=\pi M^2T_2\tilde U_2\,,\\
B&=\pi\left(\frac{T_2}{\tilde U_2}+\frac{\tilde U_2}{T_2}\right)\,,\\
\alpha&=-T_2^2 M^2 \tilde N_{(1)}^i d_{ij}\tilde N_{(1)}^j\,,\\
\beta&=2iT_2 M\,\frac{T_2+\tilde U_2}{\tilde U_2}\,\tilde N_{(1)}^i d_{ij}\tilde N_{(2)}^j-\frac{T_2}{2\pi \tilde U_2}\,\tilde N_{(2)}^i d_{ij}\tilde N_{(2)}^j\,,\\
\delta&=\left(\frac{T_2+\tilde U_2}{\tilde U_2}\right)^2\tilde N_{(2)}^i d_{ij}\tilde N_{(2)}^j\,,
\end{align}
\end{subequations}

where $\tilde N_{(1)} := {\rm Re} (\tilde N)$ and $\tilde N_{(2)} := {\rm Im}\, (\tilde N)$.

One then plugs \cref{result} into \cref{eq:intWithInsertion2}, and splits the later into its zero, positive and negative frequency parts:

\begin{equation}
\mathcal{I}_f(s)=\mathcal{I}_f^{(-)}(s)+\mathcal{I}_f^{(0)}(s)+\mathcal{I}_f^{(+)}(s)\,.
\end{equation}

\subsubsection*{Zero-frequency mode}

One has explicitely for the zero mode part of the Fourier expansion:
\begin{equation}
\mathcal{I}_f^{(0)}(s)=2\sum_{\gamma\in\Gamma_\infty\backslash\Gamma}\sqrt{T_2\tilde U_2}
\int_0^\infty\frac{\text{d}t}{t^{3/2}}\,\mathcal{M}_{s,-2}(-t)\left(\frac{\beta^{(0)}}{t}+\delta\right)\exp\left(-Bt\right)\,,
\end{equation}
which we can rewrite, using the results above, as:
\begin{equation}
\begin{split}
\mathcal{I}_f^{(0)}(s)=&\ 2^{4s+1} \,\pi^{5/2} \, \varGamma (s-\tfrac{3}{2}) \sum_{\gamma\in\Gamma_\infty\backslash\Gamma}\sqrt{T_2\tilde U_2}\sum_{\ell =0}^2 (-1)^\ell \frac{(3-\ell)_\ell \, (2s-3)_\ell}{(2s)_\ell \, \ell!} \, (s+\ell -\tfrac{3}{2})\times\\
&\times\left(-\beta^{(0)}\frac{\partial}{\partial B}+\delta\frac{\partial^2}{\partial B^2}\right)\int_0^\infty \frac{\text{d}t}{t}\, \, I_{s+\ell -\frac{3}{2}} (2 \pi t )\, e^{ - B t}\,.
\end{split}
\end{equation}
One can obtain a very explicit expression in the form:
\begin{equation}
\begin{split}
\mathcal{I}_f^{(0)}(s)=&\ 2^{4s+1} \,\pi^{5/2} \, \varGamma (s-\tfrac{3}{2}) \sum_{\gamma\in\Gamma_\infty\backslash\Gamma}\sqrt{T_2\tilde U_2}\sum_{\ell =0}^2 (-1)^\ell \frac{(3-\ell)_\ell \, (2s-3)_\ell}{(2s)_\ell \, \ell!} \, (s+\ell -\tfrac{3}{2})\times\\
&\times\left(\beta^{(0)}F_{1,s+l-\frac{3}{2}}^{(0)}(B,2\pi)+\delta F_{2,s+l-\frac{3}{2}}^{(0)}(B,2\pi)\right)\,,
\end{split}
\end{equation}
where the functions $F_{n,\nu}^{(0)}$ are defined and computed in \cref{app:functionFn}.

\subsubsection*{Positive frequency modes} 

Let us now consider the positive frequency part, the negative part being obtained from it by complex conjugation. The contribution of positive modes reads:
\begin{equation}
\begin{split}
\mathcal{I}_f^{(+)}(s)=2\sum_{M>0}\sum_{\gamma\in\Gamma_\infty\backslash\Gamma}&\sqrt{T_2\tilde U_2}e^{2i\pi M(T_1-\tilde{U_1})}\times\\
&\times\int_0^\infty\frac{\text{d}\tau_2}{\tau_2^{3/2}}\,\mathcal{M}_{s,-2}(-\tau_2)\left(\frac{\alpha}{\tau_2^2}+\frac{\beta}{\tau_2}+\delta\right)\exp\left(-\frac{A}{\tau_2}-B\tau_2\right)\,.
\end{split}
\end{equation}
One can again rewrite it as:
\begin{equation}
\begin{split}
\mathcal{I}_f^{(+)}(s)=&\ 2^{4s+1} \,\pi^{5/2} \, \varGamma (s-\tfrac{3}{2}) \sum_{M>0}\sum_{\gamma\in\Gamma_\infty\backslash\Gamma}\sqrt{T_2\tilde U_2}\sum_{\ell =0}^2 (-1)^\ell \frac{(3-\ell)_\ell \, (2s-3)_\ell}{(2s)_\ell \, \ell!} \times\\
&\times(s+\ell -\tfrac{3}{2})\left(\alpha -\beta\frac{\partial}{\partial B} + \delta \, \frac{\partial^2}{\partial B^2}\right)\int_0^\infty \frac{\text{d}t}{t}\, \, I_{s+\ell -\frac{3}{2}} (2 \pi t )\, e^{-A/t - B t}\, , 
\end{split}
\end{equation}
leading to the expression:
\begin{equation}
\begin{split}
&\mathcal{I}_f^{(+)}(s)=\ 2^{4s+1} \,\pi^{5/2} \, \varGamma (s-\tfrac{3}{2}) \sum_{M>0}\sum_{\gamma\in\Gamma_\infty\backslash\Gamma}\sqrt{T_2\tilde U_2}\sum_{\ell =0}^2 (-1)^\ell \frac{(3-\ell)_\ell \, (2s-3)_\ell}{(2s)_\ell \, \ell!} \times\\
&\times (s+\ell -\tfrac{3}{2})\left(\alpha F_{0,s+l-\frac{3}{2}}(A,B,2\pi)+\beta F_{1,s+l-\frac{3}{2}}(A,B,2\pi)+\delta F_{2,s+l-\frac{3}{2}}(A,B,2\pi)\right)\,,
\end{split}
\end{equation}
where the functions $F_n(A,B,C)$ are defined in \cref{app:functionFn}, and depend on the coeffients $A,B$ and $C$ mainly through $ u_\pm = \sqrt{A} (\sqrt{B+C} \pm \sqrt{B-C})$.

Putting all pieces together, one has the following compact expressions for the Fourier expansion of the three threshold corrections:

\paragraph{Gravitational threshold corrections:}
\begin{equation}
\label{eq:grav}
\begin{split}
24\Lambda_\text{grav}=&-\frac{n}{12}\left(\mathcal{I}(2)-5\mathcal{I}(1)-144\, \mathcal{I}_\textsc{dkl}\right)\\
&+\frac{n-24}{12}\left(\frac{1}{5}\mathcal{I}(3)-4\mathcal{I}(2)+13\mathcal{I}(1)+144\, \mathcal{I}_\textsc{dkl}\right)\\
&-\frac{1}{2}\left(\frac{1}{40}\mathcal{I}_f(3)-\frac{1}{3}\mathcal{I}_f(2)\right)\, .
\end{split}
\end{equation}

\paragraph{$E_8$ threshold corrections:}
\begin{equation}
\label{eq:E8}
\begin{split}
24\Lambda_{E_8}=&-\frac{n-12}{6}\left(\mathcal{I}(2)-5\mathcal{I}(1)-144\, \mathcal{I}_\textsc{dkl}\right)\\
&+\frac{n-24}{12}\left(\frac{1}{5}\mathcal{I}(3)-4\mathcal{I}(2)+13\mathcal{I}(1)+144\, \mathcal{I}_\textsc{dkl}\right)\\
&+\frac{n}{12}\left(\mathcal{I}(1)-1008\, \mathcal{I}_\textsc{dkl}\right)-\frac{1}{2}\left(\frac{1}{40}\mathcal{I}_f(3)-\frac{1}{2}\mathcal{I}_f(2)\right)\, .
\end{split}
\end{equation}

\paragraph{$E_7$ threshold corrections:}
\begin{equation}
\label{eq:E7}
\begin{split}
24\Lambda_{E_7}=&-\frac{n-12}{6}\left(\mathcal{I}(2)-5\mathcal{I}(1)-144\, \mathcal{I}_\textsc{dkl}\right)\\
&+\frac{n-24}{12}\left(\frac{1}{5}\mathcal{I}(3)-4\mathcal{I}(2)+13\mathcal{I}(1)+144\, \mathcal{I}_\textsc{dkl}\right)\\
&+\frac{n}{12}\left(\mathcal{I}(1)+720\, \mathcal{I}_\textsc{dkl}\right)-\frac{1}{2}\left(\frac{1}{40}\mathcal{I}_f(3)-\frac{1}{2}\mathcal{I}_f(2)\right)\, .
\end{split}
\end{equation}

\subsection{A simple subclass of models}
\label{sec:SimpleModel}

The Fourier series expansion that we have obtained above is not easy to analyse, in particular because 
the two-torus metric and the intersection form on the base $d_{ij} =\int_\mathcal{S} \omega_i \wedge \omega_j$ are intertwined 
in a non trivial way in the momentum insertion $\int\langle p_\textsc{l},p_\omega\rangle^2$. In order to unveil the role of the 
worldsheet instantons, we consider below a subclass of models that, although not really special from the physical point of view, 
allow to present the results in a much simpler way.  

Noticing that the interpretation in terms of worldsheet instantons does not depend on the precise moduli of the torus fiber, 
let us consider for convenience examples in which the momentum insertion 
$\int\langle p_\textsc{l},p_\omega\rangle^2$ is proportional to $\langle p_\textsc{l},p_\textsc{l}\rangle:=p_\textsc{l}^2$, 
namely the case where:
\begin{equation}
\label{eq:proportionality}
f(p_\textsc{l},\omega)=(n-24)\left(p_\textsc{l}^2-\frac{1}{2\pi\tau_2}\right)\, ,
\end{equation}
where the proportionality constant in front of the $p_\textsc{l}^2$ term is fixed by modularity, and where one made use of the tadpole condition~\cref{eq:tadpole}. 
It amounts to a particular relation between the torus metric and the intersection form $d_{ij}$, 
see appendix~\ref{app:mom_inser}.

For definiteness let us consider the gravitational threshold corrections corresponding to such a setting. As discussed previously, it is written:
\begin{align}
\label{eq:gravThresh}
 24\Lambda_\text{grav}=&\int_{\mathcal{F}}\text{d}\nu\, \tau_2\, 
\sum_{\mu,p_\textsc{l},p_\textsc{r}}q^{\frac{1}{2}p_\textsc{l}^2}
\bar{q}^{\frac{1}{2}p_\textsc{r}^2}\times\nonumber\\
&\times\left\{-\frac{n}{12}\frac{\hat{E}_2 E_4 E_6}{\Delta}+\frac{n-24}{12}\frac{\hat{E}_2^2 E_4^2}{\Delta}-\frac{f(p_\textsc{l},\omega)}{2}\frac{\hat{E}_2 E_4^2}{\Delta}\right\}\, ,
\end{align}
with $\text{d}\nu=\text{d}\tau_1\text{d}\tau_2/\tau_2^2$ the modular invariant measure.
Let us focus on the last term and exploit \cref{eq:proportionality}. Once again, we denote 
by  $\varLambda_{2,2}$ the partition function associated with the Narain lattice 
$\Gamma_{2,2}(T,U)$, including a factor $\tau_2$ making it modular invariant by itself. As a preliminary step, notice that:
\begin{equation}
(n-24)(D_0\varLambda_{2,2})=-\tau_2\, \sum_{\mu,p_\textsc{l},p_\textsc{r}}f(p_\textsc{l},\omega)\, q^{\frac{1}{2}p_\textsc{l}^2}
\bar{q}^{\frac{1}{2}p_\textsc{r}^2}\, ,
\end{equation}
with $D_0$ the modular covariant derivative as defined in \cref{eq:modDev}.
This allows to reexpress the last term in \cref{eq:gravThresh} simply as:
\begin{equation}
\frac{n-24}{2}\int\text{d}\nu\,(D_0\varLambda_{2,2})\frac{\hat E_2E_4^2}{\Delta}\, .
\end{equation}
Using \cref{eq:Leibniz}, an integration by part then leads to:
\begin{equation}
-\frac{n-24}{2}\int\text{d}\nu\,\varLambda_{2,2}D_{-2}\left(\frac{\hat E_2E_4^2}{\Delta}\right)\, .
\end{equation}
Exploiting again \cref{eq:Leibniz}, one computes:
\begin{equation}
D_{-2}\left(\frac{\hat E_2E_4^2}{\Delta}\right)=\frac{1}{6}\frac{E_4^3}{\Delta}+\frac{4}{3}\frac{\hat E_2E_4E_6}{\Delta}+\frac{1}{2}\frac{\hat E_2^2E_4^2}{\Delta}\, .
\end{equation}
Plugging this result into \cref{eq:gravThresh}, one obtains finally:
\begin{equation}
24\Lambda_\text{grav}=\int_{\mathcal{F}}\text{d}\nu\,\varLambda_{2,2}\left(-\frac{3n-64}{4}\frac{\hat E_2E_4E_6}{\Delta}-\frac{n-24}{6}\frac{\hat E_2^2E_4^2}{\Delta}-\frac{n-24}{12}\frac{E_4^3}{\Delta}\right)\,.
\end{equation}
Using the decompositions in terms of Niebur-Poincar\'e series, one finally obtains:
\begin{equation}
\begin{split}
24\Lambda_\text{grav}=&\,\int_{\mathcal{F}}\text{d}\nu\,\varLambda_{2,2}\left(-\frac{n-24}{30}\mathcal{F}(3,1,0)-\frac{n}{12}\mathcal{F}(2,1,0)+\right.\\
&\left.+\frac{3n-52}{2}\mathcal{F}(1,1,0)+24(n-12)\right)\,,
\end{split}
\end{equation}
which can be written in terms of \cref{eq:FourierIntegral} as:
\begin{equation}
24\Lambda_\text{grav}=-\frac{n-24}{30}\,\mathcal{I}(3)-\frac{n}{12}\,\mathcal{I}(2)+\frac{3n-52}{2}\,\mathcal{I}(1)+24(n-12)\, \mathcal{I}_\textsc{dkl}\,.
\end{equation}
Let us split the result into positive, negative and zero-frequency parts:
\begin{equation}
\Lambda_\text{grav}=\Lambda_\text{grav}^{(-)}+\Lambda_\text{grav}^{(0)}+\Lambda_\text{grav}^{(+)}\,, 
\end{equation}
that will be given separately below. 

\subsubsection*{Zero-frequency mode}

It turns out that one can have a very explicit expression for the zero mode part of the above expression in terms of real analytic Eisenstein series, defined by:
\begin{equation}
E(z,\rho):=\frac{1}{2}\sum_{(m,n)=1}\frac{\text{Im}(z)^\rho}{|m+zn|^{2\rho}}\,.
\end{equation}

The zero-frequency mode of the gravitational threshold correction in the above example is then given by:
\begin{equation}
\begin{split}
\Lambda_\text{grav}^{(0)}=&\,\frac{\pi}{90T_2^2}\Big\{15(3n-52)T_2^2\,E(U,1)-5nT_2\,E(U,2)-12(n-24)\,E(U,3)\Big\}\\
&+(n-12)\, \mathcal{I}_\textsc{dkl}\,.
\end{split}
\end{equation}
In the $K3\times T^2$ case, it reduces to:
\begin{equation}
\Lambda_\text{grav}^{(0)}=\,\frac{2\pi}{3}\Big\{5\,E(U,1)-2T_2^{-1}\,E(U,2)\Big\}+12\, \mathcal{I}_\textsc{dkl}\,.
\end{equation}

\subsubsection*{Positive frequency part}

The positive frequency part can also be written explicitely in terms of the Niebur-Poincar\'e series themselves, cf. \cref{eq:intWithoutInsertion}:
\begin{equation}
\begin{split}
\Lambda_\text{grav}^{(+)}=&\,\frac{1}{30\times 24}\sum_{M>0}\frac{e^{2i\pi MT_1}}{M}\Big\{30(3n-52)\,\mathcal{W}_{1,0}(MT_2)\mathcal{F}(1,M,0;U)\\
&-5n\,\mathcal{W}_{2,0}(MT_2)\mathcal{F}(2,M,0;U)
-2(n-24)\,\mathcal{W}_{3,0}(MT_2)\mathcal{F}(3,M,0;U)\Big\}\, ,
\end{split}
\end{equation}
which reduces for  $K3\times T^2$ to:
\begin{equation}
\Lambda_\text{grav}^{(+)}=\,\frac{1}{6}\sum_{M>0}\frac{e^{2i\pi MT_1}}{M}\Big\{5\,\mathcal{W}_{1,0}(MT_2)\mathcal{F}(1,M,0;U)-\mathcal{W}_{2,0}(MT_2)\mathcal{F}(2,M,0;U)\Big\}\,.
\end{equation}
Given that $\mathcal{W}_{1+\ell,0}(MT_2) \sim (MT_2)^{-\ell} e^{-2\pi M T_2} 
\times (\text{polynomial in } M T_2)$, one has in both cases a sum over $M \in \mathbbm{Z}_{>0}$ which 
represents the sum over the wrapping number of a worldsheet instanton around 
the two-torus fiber of the principal bundle $T^2\hookrightarrow\mathcal{M}\stackrel{\pi}{\to}\mathcal{S}$, which is of volume $T_2$.

Even though for $n<24$ the torus fiber is only a torsional two-cycle, it appears that worldsheet instantons, corresponding to holomorphic maps from the heterotic worldsheet to $\mathcal M$ wrapping the fiber, do contribute to the threshold corrections, 
for any wrapping number. 

Would we have decided to work with the $\text{Spin}(32)/\mathbbm Z_2$ heterotic string, this discussion should also be considered in the context of type I flux compactifications via S-duality~\cite{Dasgupta:1999ss}. Then, the heterotic thresholds encompass both the spacetime perturbative and non-perturbative effects on the type I side, the latter corresponding to Euclidean D1-branes wrapping the torus fiber.

\section{Conclusion}

In this work, we have computed the one-loop threshold corrections to the gauge and gravitational couplings in a 
large class of $\mathcal{N}=2$ heterotic compactifications on non-K\"ahler manifolds with three-form flux. We have obtained 
the result from the new supersymmetric index that was computed in~\cite{Israel:2015aea,Israel:2016xfu}.

The results were first given in terms of Niebur-Poincar\'e series, exhibiting invariance under perturbative 
$O(2,2;\mathbbm{Z})$ dualities, and second as a Fourier series expansion, allowing to isolate the contributions of the 
worldsheet instantons wrapping the two-torus fiber of the principal bundle. 

By S-duality our results apply to  D-instanton corrections in some Ramond-Ramond backgrounds. A better understanding 
of the physics behind these instanton corrections would involve then studying D1-instanton probes in these flux backgrounds 
of type I supergravity. We plan to come back to this problem in the near future.\footnote{In~\cite{Kim:2006qs} heterotic five-branes wrapping the torus fiber have been studied. However the  physics is 
not the same because the coupling to the NS-NS flux is different.}

A generalization of our results to models with Abelian gauge bundle over the total space is also worthwile considering, given that 
the new supersymmetric index has also been computed in those cases~\cite{Israel:2016xfu}. These examples are especially important from the four-dimensional effective field-theory perspective, as the threshold corrections will then be functions of the bundle moduli, while the torus moduli are frozen to discrete values for a generic choice of torus principal bundle. 

It would be very interesting to consider compactifications with torsion with reduced or without 
supersymmetry, that can be obtained as freely orbifolds of the $\mathcal{N}=2$ models~\cite{2008arXiv0807.0827B} and 
investigate whether, as for $K3\times T^2$ models~\cite{Angelantonj:2015nfa}, one obtains a universal behavior. 

\subsection*{Acknowledgements}

We thank Nick Halmagyi, Ruben Minasian, Ilarion Melnikov, Ronen Plesser and Jan Troost, 
for discussions and correspondence. This work was conducted within the 
ILP LABEX (ANR-10-LABX-63) supported by French state funds managed by the 
ANR (ANR-11-IDEX-0004-02) and by the project QHNS in the program ANR Blanc 
SIMI5 of Agence National de la Recherche. 


\appendix

\section{Theta functions, modular covariant derivative and hypergeometric functions}
\label{app:hypergeom}

We define the odd Jacobi theta function and the Dedekind eta function by the following infinite products:
\begin{subequations}
\begin{align}
\theta(\tau,z)&:=-i\,q^{\frac{1}{8}}y^{\frac{1}{2}}\prod_{n=1}^{\infty}\left(1-q^n\right)\left(1-yq^n\right)\left(1-y^{-1}q^{n-1}\right)\,.\\
\eta(\tau)&:=q^{\frac{1}{24}}\prod_{n=1}^{\infty}\left(1-q^{n}\right)\, ,
\end{align}
\end{subequations}
with $q:=\exp(2i\pi\tau)$ and $y:=\exp(2i\pi z)$.
The discriminant modular form is given in terms of the Dedekind eta function by:
\begin{equation}
\Delta(\tau):=\eta(\tau)^{24}\,.
\end{equation}
Given an even integral lattice $\Gamma$, whose pairing we denote:
\begin{equation}
\langle\cdot,\cdot\rangle:\Gamma\times\Gamma\rightarrow\mathbbm{Z}\,,
\end{equation}
and an element $\mu\in\Gamma^\star/\Gamma$ in its discriminant group, 
we define its associated theta-function with characteristic $\mu$ as a refined generated function:
\begin{equation}
\begin{split}
\Theta_\mu^\Gamma:\mathbbm H\times (\Gamma\otimes\mathbbm C)&\rightarrow \mathbbm C\\
(\tau,z)&\mapsto\sum_{v\in\Gamma+\mu}e^{i\pi\left(\langle v,v\rangle\tau+2\langle v,z\rangle\right)}\,.
\end{split}
\end{equation}
Let us recall the definition of the $SL_2(\mathbbm Z)$ normalized Eisenstein series of weight $2w$:
\begin{equation}
E_{2w}(\tau):=\frac{1}{2\zeta(2w)}\sum_{(m,n)\in (\mathbbm Z^{*})^2}\frac{1}{|m+\tau n|^{2w}}\,.
\end{equation}
We define the following weight-$2$ modular covariant derivative acting on the space of weight $w$ modular forms:
\begin{equation}
\label{eq:modDev}
\begin{split}
D_w:\ \mathcal M_w&\rightarrow\mathcal M_{w+2}\\
f&\mapsto\left(\frac{i}{\pi}\frac{\partial}{\partial\tau}+\frac{w}{2\pi\tau_2}\right)f\, .
\end{split}
\end{equation}
Notice that this modular covariant derivative satisfies the Leibniz rule:
\begin{equation}
\label{eq:Leibniz}
D_{w+r}(\psi_w\phi_r)=(D_w\psi_w)\phi_r+\psi_wD_r(\phi_r)\, .
\end{equation}
We give two identities due to Ramanujan involving the Eisenstein series:
\begin{subequations}
\begin{align}
D_4E_4&=\frac{2}{3}\left(E_6-\hat E_2 E_4\right)\, ,\\
D_6E_6&=E_4^2-\hat E_2 E_6
\end{align}
\end{subequations}
The confluent hypergeometric function ${}_1 F _1\left(a;c;z\right)$ is defined by:
\begin{equation}
{}_1 F _1\left(a;c;z\right):=\sum_{n=0}^{\infty}\frac{(a)_n}{(c)_n}\frac{z^n}{n!}\,,
\end{equation}
with $(q)_n$ the Pochhammer symbol, or rising factorial.
The hypergeometric function ${}_2 F _1\left(a,b;c;z\right)$ is defined by:
\begin{equation}
{}_2 F _1\left(a,b;c;z\right):=\sum_{n=0}^{\infty}\frac{(a)_n (b)_n}{(c)_n}\frac{z^n}{n!}\,.
\end{equation}
We give the expression of the hypergeometric function ${}_2 F _1$ for some specific values of its arguments:
\begin{subequations}
\label{eq:Hypergeom_Simple}
\begin{align}
{}_2 F _1(2,4,6,t^{-1})&=-\frac{10}{3}  t^2 \left(24 t^2+6 (4 t-3) t^2 \log \left(\frac{t-1}{t}\right)-6 t-1\right)\,,\\
{}_2 F _1(1,3,4,t^{-1})&=-\frac{3}{2} t \left(2 t^2 \log \left(\frac{t-1}{t}\right)+2 t+1\right)\,,\\
{}_2 F _1(2,2,4,t^{-1})&=-6 t^2 \left((2 t-1) \log \left(\frac{t-1}{t}\right)+2\right)\,,\\
{}_2 F _1(1,1,2,t^{-1})&=-t \log \left(\frac{t-1}{t}\right)\,,\\
{}_2 F _1(2,3,6,t^{-1})&=10 t^2 \left(12 t^2+6 \left(2 t^2-3 t+1\right) t \log \left(\frac{t-1}{t}\right)-12 t+1\right)\,,\\
{}_2 F _1(1,2,4,t^{-1})&=3 t \left(2 t+2 (t-1) t \log \left(\frac{t-1}{t}\right)-1\right)\,,\\
{}_2 F _1(3,3,6,t^{-1})&=-30 t^3 \left(\left(6 t^2-6 t+1\right) \log \left(\frac{t-1}{t}\right)+6 t-3\right)\,.
\end{align}
\end{subequations}

\section{Coefficients}
\label{app:coeff}

In this appendix, we will determine the exact coefficients entering in the computation of the Fourier expansion representation of the integral \cref{eq:intWithInsertion}: 
\begin{equation}
\mathcal{I}_f(s):=\int_\cF \text{d}\nu \, \cF (s,1,-2)\, \tau_2\sum_{\mu,p_\textsc{l},p_\textsc{r}}q^{\frac{1}{2}p_\textsc{l}^2}\bar q^{\frac{1}{2}p_\textsc{r}^2}
f(p_\textsc{l},\omega)\,,
\end{equation}
with quadratic momentum insertion:
\begin{equation}
f(p_\textsc{l},\omega) = \tilde d_{ij} p^i_\textsc{l} p^j_\textsc{l} -\frac{n-24}{2\pi\tau_2}\,,
\end{equation}
with the metric $\tilde d$ defined in \cref{app:mom_inser}.

As mentioned in \cref{sec:FourierSeries}, the first step in deriving the Fourier representation is to first perform the integral over the Lagrange multiplier $\tau_1$ to impose the constraint on the lattice momenta, solve explicitely the constaint, and perform a suitable Poisson resummation.

Let us now introduce some notation for the lattice. First, the $\Gamma_{2,2}(T,U)$ Narain lattice elements can be written in a complex basis as 
\begin{align*}
p_L &= \frac{1}{\sqrt{U_2 T_2}} \left(m_2 - U m_1 + \bar{T} (n^1 + U n^2) \right)\\
p_R &= \frac{1}{\sqrt{U_2 T_2}} \left(m_2 - U m_1 + T (n^1 + U n^2) \right)
\end{align*}
In complex notation the scalar product becomes 
\begin{equation}
\langle p_L,p_L ' \rangle  =\frac{1}{2}\left( |p_L + p_L ' |^2 - |p_L|^2 - |p_R|^2 \right) 
= \text{Re}\, (p_L \bar{p}_L ')\, .
\end{equation}

The BPS constraint $\frac{1}{4}(|p_L|^2 - |p_R|^2) = m_1 n^1 + m_2 n^2 = 1$ is solved, for coprime $(n^1,n^2)$, as 
\begin{equation}
m_1 = m_1^\star + \tilde{M} n^2 \ , \quad
m_2 = m_2^\star - \tilde{M} n^1 \, ,
\label{eq:dec}
\end{equation}
where $m_1^\star$ is a modular inverse of $n^1$ modulo $n^2$, and  $m_2^\star$ a modular inverse of $n^2$ modulo $n^1$.

As mentioned in \cref{sec:FourierSeries}, after solving the constraint on momenta as above, one ends up with an expression of the following form:
\begin{equation}
\sum_{\tilde M \in \mathbbm{Z}} e^{-\pi a \tilde{M}^2 +2i\pi b \tilde{M}} \left(c \tilde{M}^2 
+ d \tilde{M} + e \right)\,,
\end{equation}
to be Poisson resummed over the variable $\tilde M$.

The Poisson resummation formula on $\mathbbm Z$:
\begin{equation}
 \sum_{n=-\infty}^{\infty}f(n)=\sum_{k=-\infty}^{\infty}\tilde f(k)\,,
\end{equation}
 gives the following general formulae:
\begin{equation}
\begin{split}
\sum_{n\in \mathbbm{Z}} e^{-\lambda n^2} &= \sqrt{\frac{\pi}{\lambda}}\, \sum_{k\in \mathbbm{Z}} e^{-\pi^2 k^2 /\lambda}\,,
\\
\sum_{n\in \mathbbm{Z}} n\, e^{-\lambda n^2} &= -i \sqrt{\frac{\pi}{\lambda}}\,  \sum_{k\in \mathbbm{Z}} \frac{\pi k}{\lambda}\, e^{-\pi^2 k^2 /\lambda}\,,
\\
\sum_{n\in \mathbbm{Z}} n^2\, e^{-\lambda n^2} &= \sqrt{\frac{\pi}{\lambda}}\, \sum_{k\in \mathbbm{Z}} \left( \frac{1}{2\lambda} - \frac{\pi^2 k^2}{\lambda^2}\right)\, e^{-\pi^2 k^2 /\lambda}\,.
\end{split}
\end{equation}
Using these results, one obtains the Poisson resummed expression:
\begin{equation}
\begin{split}
&\sum_{\tilde M \in \mathbbm{Z}} e^{-\pi a \tilde{M}^2 +2i\pi b \tilde{M}} \left(c \tilde{M}^2 
+ d \tilde{M} + e \right) \\
&= \frac{1}{\sqrt{a}} \sum_{M \in \mathbbm{Z}} 
e^{-\frac{\pi}{a}(M-b)^2} \left\{ \left(\frac{1}{2\pi a}  - \left(\frac{b-M}{a} \right)^2 \right) c
+ \frac{i(b-M)}{a} d + e \right\}\\
&= \frac{1}{\sqrt{a}} \sum_{M \in \mathbbm{Z}} 
e^{-\frac{\pi}{a}(M-b)^2} \left\{ \left(\frac{1}{2\pi a} c - \frac{b^2}{a^2}c + \frac{ib}{a} d +e \right)
+ \frac{1}{a}\left( \frac{2b}{a}c - i d \right) M
- \frac{c}{a^2} M^2  
\right\}\,.
\end{split}
\end{equation}
In order to determine the various coefficients $(a,b,c,d,e$), let us expand the left momentum as 
\begin{equation}
p_L = \frac{1}{\sqrt{U_2 T_2}} \left( \underbrace{m_2^\star - U m_1^\star + \bar{T} (n^1 + U n^2)}_{P_L^\star}
-\tilde{M} \underbrace{(n^1 + U n^2)}_{\hat{P}}\right)
\end{equation}
In the following one will consider the $SL(2;\mathbbm{Z})$ transformation related to the solution of the BPS constraint:
\begin{equation}
\tilde{U} = \frac{m_1^\star U - m_2^\star}{n^1 + n^2 U}\,,
\end{equation}
implying in particular that:
\begin{equation}
\tilde{U}_2 = \frac{U_2}{|n^1 + n^2 U|^2} = \frac{U_2}{|\hat{P}|^2}\,.
\end{equation}
We remark also that:
\begin{equation}
\frac{P_L^\star}{\hat{P}} = \bar{T} - \tilde{U} 
\end{equation}
Now we consider the vector of two-forms that appears in the insertion. Considering a basis $\{ \varpi_\ell \}$ of $\text{Pic} (\mathcal{S})$, we 
expand, in complex notation 
\begin{equation}
p_w = \sqrt{\frac{T_2}{U_2}}\underbrace{\left(N^\ell_1 + U N^\ell_2\right)}_{N^\ell} \varpi_\ell \, ,
\end{equation}
and introduce the intersection form $d_{\ell k} = \int \varpi_\ell \wedge \varpi_k$. Notice that $(N^\ell_1,N^\ell_2)$ transforms 
as a doublet under $SL(2;\mathbbm{Z})_U$. 
From this one can compute the scalar products that appear in the insertion. One obtains for the quadratic momentum insertion:
\begin{equation}
\begin{split}
f(p_\textsc{l},\omega)=&\,\int_\mathcal{S} \langle p_\omega,p_\textsc{l} \rangle^2 - \frac{1}{4\pi \tau_2}\langle p_\omega,p_\omega \rangle \\ 
=&\,\frac{1}{\tilde{U}_2^2} \Big\{
 d_{\ell k}\, \text{Re}\, \left(\tilde{N}^\ell \left(T-\bar{\tilde{U}} - \tilde{M} \right)\right) 
 \, \text{Re}\,\left(\tilde{N}^k \left(T-\bar{\tilde{U}} - \tilde{M} \right)\right)\\ 
&- \frac{ T_2\tilde{U}_2}{4\pi \tau_2} 
d_{\ell k} \, \text{Re}\, \left(\tilde{N}^\ell \overline{\tilde{N}^k}\right)\Big\}
\end{split}
\end{equation}
Out of this expression one can first collect the term in $\tilde{M}^2$, namely:
\begin{equation}
\label{eq:alpha}
c = \frac{d_{\ell k}}{\tilde{U}_2^2} \text{Re}\, \left(\tilde{N}^\ell\right) \text{Re}\, \left(\tilde{N}^k\right)\,,
\end{equation}
then term linear in $\tilde{M}$, which reads (using the symmetry of the intersection form):
\begin{equation}
\label{eq:beta}
d = - \frac{2d_{\ell k}}{\tilde{U}_2^2} \left[(T_1 - \tilde{U}_1)\, \text{Re}\, \left(\tilde{N}^\ell\right) 
- (T_2 + \tilde{U}_2 )\, \text{Im}\, \left(\tilde{N}^\ell\right) \right]\text{Re}\, \left(\tilde{N}^k\right)\,,
\end{equation}
and finally the constant term given by:
\begin{equation}
\label{eq:gamma}
\begin{split}
e =&\ \frac{d_{\ell k}}{\tilde{U}_2^2}
 \left[(T_1 - \tilde{U}_1)\, \text{Re}\, \left(\tilde{N}^\ell\right) 
- (T_2 + \tilde{U}_2 )\, \text{Im}\, \left(\tilde{N}^\ell\right) \right]\times\\
&\times \,\left[(T_1 - \tilde{U}_1)\, \text{Re}\, \left(\tilde{N}^k\right) 
- (T_2 + \tilde{U}_2 )\, \text{Im}\, \left(\tilde{N}^k\right) \right]\\ 
&- \frac{T_2}{4\pi \tau_2 \tilde{U}_2} 
d_{\ell k} \, \text{Re}\, \left(\tilde{N}^\ell \overline{\tilde{N}^k}\right)\,.
\end{split}
\end{equation}
We are now ready to consider the Poisson resummation of the result, organised in powers of the dual variable $M$. In the 
exponential, we have:
\begin{equation}
\begin{split}
\exp\left(-\frac{\pi\tau_2}{2}(|p_L|^2 + |p_R|^2 ) \right)  &= 
 \exp\left(-\frac{\pi \tau_2}{U_2 T_2} |P_L^\star - \tilde{M} \hat{P}|^2 + 2\pi \tau_2\right)\\
 & = 
\exp\left(-\frac{\pi \tau_2}{\tilde{U}_2 T_2} |\bar{T}-\tilde{U} - \tilde{M}|^2 + 2\pi \tau_2\right)\\
 &= \exp\left(2\pi \tau_2 -   \frac{\pi \tau_2}{\tilde{U}_2 T_2}|\bar{T}-\tilde{U}|^2\right)\times\\
&\ \ \ \, \times\exp\left(-\frac{\pi \tau_2}{\tilde{U}_2 T_2} \tilde{M}^2 + 2i\pi  \tilde{M}\frac{i\tau_2}{\tilde{U}_2 T_2}
( \tilde{U}_1-T_1)\right)\,,
\end{split}
\end{equation}
from which one can once again extract the coefficient of the quadratic and linear terms in $\tilde M$ to obtain $a$ and $b$. 
Adding up the terms quadratic, linear and constant in the dual dummy variable $M$, one can finally read up the summand of the Poisson resummed momentum insertion:
\begin{equation}
\begin{split}
&\sqrt{\frac{\tilde U_2 T_2}{\tau_2}}\sum_{M\in \mathbbm Z} \Bigg\{- \frac{T_2^2}{\tau_2^2}\, 
d_{\ell k} 
\left[\text{Re}\, \left(\tilde{N}^\ell\right)M  
-i \frac{\tau_2}{\tilde{U}_2 T_2}(T_2 + \tilde{U}_2 )\, \text{Im}\, \left(\tilde{N}^\ell\right)\right]\times\\
&\times\left[\text{Re}\, \left(\tilde{N}^k\right) M 
-i \frac{\tau_2}{\tilde{U}_2 T_2}(T_2 + \tilde{U}_2 )\, \text{Im}\, \left(\tilde{N}^k\right)\right]- \frac{T_2}{2\pi \tau_2 \tilde{U}_2} d_{\ell k}\, \text{Im}\, \left(\tilde{N}^\ell\right)\text{Im}\, \left(\tilde{N}^k\right)\Bigg\}\times\\
&\times \exp\left\{-\frac{\pi\tilde U_2 T_2}{\tau_2}\left(M-\frac{i\tau_2}{\tilde U_2 T_2}(\tilde U_1-T_1)\right)^2+2\pi\tau_2-\frac{\pi\tau_2}{\tilde U_2 T_2}|\bar T-\tilde U|^2\right\}\,,
\end{split}
\end{equation}
to be used in \cref{sec:Fourierseries}.

\section{Relevant functions for the Fourier series representation}
\label{app:functionFn}

In this appendix, we define various functions defined by an integral involving modified Bessel functions, and relevant for the computation of the Fourier series representation of the various threshold corrections in \cref{sec:Fourierseries}. To obtain the expressions below, one extensively makes use of the following Bessel functions identity:
\begin{equation}
2 \frac{d}{dx} C_\alpha (x)=C_{\alpha-1} (x) + C_{\alpha +1} (x)\,,
\end{equation}
where $C_\alpha$ denotes $I_\alpha$ or $e^{i\pi \alpha} K_\alpha$. 
\paragraph{Zero-frequency mode:}
Let us first define $\forall (B,C,\nu,n)\in\mathbbm C\times\mathbbm R\times\mathbbm C\times\mathbbm N$ such that $\text{Re}(B)\geqslant C$:
\begin{equation}
F_{n,\nu}^{(0)}(B,C):=(-1)^n\frac{\partial^n}{\partial B^n}\int_0^\infty\frac{\text{d}t}{t}\,I_{\nu}(C t)\,e^{-Bt}\,,
\end{equation}
relevant for the computation of the zero mode component of the Fourier series expansion. Following Erdelyi, one can compute $F_{n,\nu}^{(0)}$ explicitely:
\begin{subequations}
\begin{align}
&F_{0,\nu}^{(0)}(B,C)=\frac{C^\nu \left(B+\sqrt{B^2-C^2}\right)^{-\nu}}{\nu}\,,\\
&F_{1,\nu}^{(0)}(B,C)=\frac{C^\nu \left(B+\sqrt{B^2-C^2}\right)^{-\nu}}{\sqrt{B^2-C^2}}\,,\\
&F_{2,\nu}^{(0)}(B,C)=\frac{C^\nu \left(B+\sqrt{B^2-C^2}\right)^{-\nu}\left(B+\nu\sqrt{B^2-C^2}\right)}{\left(B^2-C^2\right)^{3/2}}\,.
\end{align}
\end{subequations}
\paragraph{Positive frequency modes:}
We also define $\forall (A,B,C,\nu,n)\in\mathbbm C^2\times\mathbbm R\times\mathbbm C\times\mathbbm N$:
\begin{equation}
F_{n,\nu}(A,B,C):=(-1)^n\frac{\partial^n}{\partial B^n}\int_0^\infty\frac{\text{d}t}{t}\,I_{\nu}(C t)\,e^{-Bt-A/t}\,,
\end{equation}
relevant for the computation of the positive frequency modes of the Fourier series expansion.
One then computes the following expressions:
\begin{subequations}
\begin{align}
&F_{0,\nu}(A,B,C)=2 I_{\nu }(u_-) K_{\nu }(u_+)\,,\\
&F_{1,\nu}(A,B,C)=\frac{4 A}{u_-^2-u_+^2}\Big(u_- I_{\nu -1}(u_-) K_{\nu }(u_+)+u_+ I_{\nu }(u_-) K_{\nu -1}(u_+)\Big)\,,\\
&F_{2,\nu}(A,B,C)=\frac{8 A^2}{u_- \left(u_-^2-u_+^2\right)^3}\Big(2 u_-^2 u_+ \left(u_-^2-u_+^2\right) I_{\nu -1}(u_-) K_{\nu -1}(u_+)\nonumber\\
&+u_- I_{\nu -2}(u_-) \left(u_-^4 K_{\nu }(u_+)-2 (\nu +1) u_-^2 u_+ K_{\nu -1}(u_+)-u_+^4 K_{\nu -2}(u_+)\right)\nonumber\\
&-2 u_+^2 I_{\nu -1}(u_-) \left((\nu +1) u_-^2-(\nu -1) u_+^2\right) K_{\nu -2}(u_+)\Big)\,,
\end{align}
\end{subequations}
where we have introduced the following convenient combinations:
\begin{equation}
u_\pm:=\sqrt{A}\left(\sqrt{B+C}\pm\sqrt{B-C}\right)\,.
\end{equation}

\section{Generic momentum insertion}
\label{app:mom_inser}

In \cref{sec:SimpleModel}, we discussed a simple class of models for which the momentum insertion takes a particularly simple form. In this short appendix, we want to understand in more detail the constraint \cref{eq:proportionality}.

The data of the compactification involves an even integral lattice $\Gamma_\textsc{l}$ naturally associated to the rational Narain lattice $\Gamma_{2,2}$. In the following we denote this lattice $\Gamma_\textsc{l}$ simply $\Gamma$.
One associates to this lattice the theta function $\Theta^\Gamma:\mathbbm H\times(\Gamma\otimes\mathbbm C)\rightarrow\mathbbm C$:
\begin{equation}
\Theta^\Gamma(\tau,z)=\sum_{v\in\Gamma}e^{i\pi\left(\langle v,v\rangle+2\langle v,z\rangle\right)}\,,
\end{equation}
whose second argument lives in the complexification of the lattice $\Gamma$. More precisely, the rational Narain lattice partition function involves such a theta function with an extra characteristic $\mu$, namely the summation vector runs over the shifted lattice $\Gamma+\mu$, where $\mu$ belongs to the discriminant group $\Gamma^\star/\Gamma$. 
In our situation, we actually have two lattices, $\Gamma$ and $\text{Pic}(\mathcal S)$. The inner product on $\Gamma$ is denoted $\langle\cdot,\cdot\rangle$, and the one on $\text{Pic}(\mathcal S)$ is defined via the composition:
\begin{equation}
(\cdot,\cdot):\Gamma\times\Gamma\rightarrow \Gamma\wedge\Gamma\xrightarrow{\int_{\mathcal S}}\mathbbm Z\,,
\end{equation}
In our situation, the second argument of the theta function actually lives in a further extension of the lattice $\Gamma$:
\begin{equation}
\Theta^\Gamma\left(\tau,\frac{p_\omega}{2i\pi}\right)=\sum_{v\in\Gamma}e^{i\pi\langle v,v\rangle+\langle v,p_\omega\rangle}\,,
\end{equation}
with $p_\omega\in\Gamma\otimes\text{Pic}(\mathcal S)\otimes\mathbbm C$. Hence, denoting $\{e_i\}$ and $\{\epsilon_a\}$ a basis of $\Gamma$ and $\text{Pic}(\mathcal S)$ respectively, we have:
\begin{subequations}
\begin{align}
p_\omega&=\omega^{ia}\,e_i\otimes\epsilon_a\,,\\
v&=v^i\,e_i\,,
\end{align}
\end{subequations}
with $\omega^{ia}\in\mathbbm C$. The matrix $(\omega^{ia})$ specifies the data of the torus fibration, and is fixed once and for all for a given model. $(\omega_{ia})$ should be viewed as connecting the two \textit{a priori} independent integral even lattices $\Gamma$ and $\text{Pic}(\mathcal S)$. In the above function,
$\langle v,p_\omega\rangle$ should be understood as:
\begin{equation}
\langle v,p_\omega\rangle:=v^ig_{ij}\omega^{ja}\,\epsilon_a\,,
\end{equation}
with $g$ the metric on the lattice $\Gamma$, namely $g_{ij}:=\langle e_i,e_j\rangle$, not to be confused with the metric on the Narain lattice $\Gamma_{2,2}$. We also define the metric $d$ on the lattice $\text{Pic}(\mathcal S)$ by:
\begin{equation}
d_{ab}:=(\epsilon_a,\epsilon_b)=\int_\mathcal S\,\epsilon_a\wedge\epsilon_b\,.
\end{equation}
Let us also define the pull-back metric:
\begin{equation}
\tilde d^{ij}:=\omega^{ia}\omega^{jb}\,d_{ab}\,.
\end{equation}
Notice that we can define a natural inner product on $\Gamma\otimes\text{Pic}(\mathcal S)$, which we denote by a dot, in the following way: given two elements 
$\alpha=\alpha^{ia}\,e_i\otimes\epsilon_a$ and $\beta=\beta^{ia}\,e_i\otimes\epsilon_a$, we define:
\begin{equation}
\alpha\cdot\beta:=g_{ij}d_{ab}\alpha^{ia}\beta^{jb}\,.
\end{equation}
Let us look at the simplified case for which $\int_\mathcal S\langle v,p_\omega\rangle^2\propto (n-24)\langle v,v\rangle$, the $(n-24)$ coefficient originating 
from the tadpole cancellation condition $p_\omega\cdot p_\omega=2(n-24)$. Let us simply express the momentum insertion in terms of the lattice data in the following way:
\begin{equation}
\begin{split}
\int_\mathcal S\langle v,p_\omega\rangle^2=&\, v^iv^k\,g_{ij}g_{kl}\tilde d^{jl}=\tilde d_{ik}v^iv^k\,.
\end{split}
\end{equation}
Therefore, we see that the insertion is proportional to $(n-24)\langle v,v\rangle$ if and only if $\tilde d\propto g$.

\newpage

\bibliographystyle{JHEP}
\bibliography{biblioJ}

\end{document}